\documentclass[reprint,onecolumn,aps,prd,nofootinbib,showpacs,notitlepage,superscriptaddress]{revtex4-2}
\usepackage{amsmath,bm}
\usepackage{hyperref}
\usepackage{enumerate,enumitem}
\usepackage{graphicx}
\usepackage{bm}
\usepackage{hyperref}
\usepackage{graphicx}
\usepackage[dvipsnames]{xcolor}
\usepackage{amsfonts}
\usepackage{amssymb,amsthm}
\usepackage{mathrsfs}
\usepackage{bbold}
\usepackage{orcidlink}

\newcommand{\ham}{\mathcal{H}}
\newcommand{\diff}{\mathcal{D}}
\newcommand{\shift}{{N^x}}

\newcommand{\erad}{E^x{}}
\newcommand{\ephi}{E^\varphi{}}
\newcommand{\krad}{K_x}
\newcommand{\kang}{K_\varphi}
\newcommand{\qbat}{\erad}
\newcommand{\qbi}{\ephi}
\newcommand{\pbat}{\krad}
\newcommand{\pbi}{\kang}
\newcommand{\pphi}{P_\phi}
\newcommand{\hor}{\mathcal{Z}}

\begin{document}

\title{Formation of nonsingular spherical black holes with holonomy corrections}

\author{Asier Alonso-Bardaji\,\orcidlink{0000-0002-8982-0237}\,}
    \email{asier.alonso@ehu.eus}
    \affiliation{Aix Marseille Univ., Univ. de Toulon, CNRS, CPT, UMR 7332, 13288 Marseille, France}

\begin{abstract}
We present a canonical model of spherical gravity with covariant corrections motivated by loop quantum gravity. The effective Hamiltonian defines univocally a family of geometries that generalizes the Lema\^itre-Tolman-Bondi spacetimes, and they can be matched to the vacuum of the theory across a timelike hypersurface comoving with the flow of matter. Such is precisely the complete spacetime picture of a spherical star subject to its own gravitational pull. The singularity gets replaced with a spacelike boundary in the trapped region of spacetime, where the curvature remains finite, and the area of the orbits of the spherical symmetry group attains its infimum. Observers falling into the black hole are doomed to travel forever towards this boundary without ever reaching it. The theory also predicts the formation of stable black-hole remnants of Planckian mass.
\end{abstract}

\maketitle

\section{Introduction}

The success of homogeneous models of loop quantum cosmology to predict the singularity resolution through a quantum bounce at Planckian densities \cite{Ashtekar_2006_1,Ashtekar_2006, Rovelli_2014, Diener_2014} fired the shot to search effective spherical black-hole models in loop quantum gravity (LQG). The first attempts tried to implement the analogy between Kantowski-Sachs homogeneous spacetimes and the interior of the Schwarzschild black hole. However, this neglects all the complexity earned when jumping to nonhomogeneous models, because, in contrast to cosmological scenarios, the spherically symmetric setup is constrained by the hypersurface deformation algebra. In other words, the so-called diffeomorphism and Hamiltonian constraints need to follow certain known commutation relations. And, further, the phase-space dependent structure functions emerging from that algebra define the metric of the spacetime manifold in a very specific way \cite{Teitelboim73,Pons_1997}. Otherwise, the same Hamiltonian produces different geometries for each gauge solution, since the different line elements are not related by a change of coordinates. In general, the usual Schwarzschild gauge (the static and diagonal coordinates) acquires no corrections. Other gauge solutions may show a singularity resolution, but there is no preferred reason to choose this geometry over the classical one. Even if some quantum fuzziness on the Planck-scale ``geometry'' could be expected, these effects are beyond the scope of this kind of studies. 

Effective black-hole models motivated by LQG have proliferated in the literature, both in vacuum \cite{Modesto_2010,Ben_Achour_2018,Gambini_2020,Kelly_2020,Ashtekar_2018,Gambini_2023}, and including simple matter fields such as irrotational dust or scalar fields \cite{Liu_2014,Ben_tez_2020,Giesel_2021,Alonso-Bardaji:2021tvy,Gambini_2022,Husain_2022,Giesel_2023,Lewandowski_2023,Fazzini:2023scu,Fazzini:2023ova,Cipriani_2024,GieselUnknown,Wilson-EwingUnknown}. We refer the interested reader to the recent review \cite{Ashtekar_2023}, and references therein. These nonsingular geometries describe the expected LQG effects of singularity resolution through a minimum positive area gap. Let us focus first on the vacuum solutions. In general, and unlike in the homogeneous case, there is an additional horizon. If the spheres attain their minimum before getting there, they become antitrapped and emerge into a parallel (time-reversed) universe, resembling the big bounce in cosmological models. If the inner horizon appears before getting to the minimum of the area-radius function, the spheres are nontrapped, and the light cones may pivot and bounce to the same asymptotic region. The dynamical formation of such black holes turns out more complex, and there is a wide variety of predictions, mainly focused on Lema\^itre-Tolman-Bondi (LTB) models or quantum Oppenheimer-Snyder collapses. 

In this work, we demand that the Hamiltonian and its associated metric are explicitly covariant. That is, each gauge solution on phase space must define a chart (and domain) of the same geometry. The search for effective models of LQG with a well-defined spacetime structure began some time ago (see \cite{Tibrewala_2014,Bojowald_2015} and references therein), and it produced the first positive results in Refs.~\cite{Alonso-Bardaji:2021yls,Alonso-Bardaji:2022ear,Alonso-Bardaji:2023niu}. These were soon generalized to the most general Hamiltonian quadratic in derivatives~\cite{Alonso-Bardaji:2023vtl,Bojowald:2023xat}. {In contrast to approaches that seek explicitly covariant Lagrangian models of LQG (see, for instance, Refs.~\cite{Han:2022rsx,Giesel:2024mps} for recent developments relating polymerized models with mimetic gravity), this framework offers an alternative guideline for identifying generally covariant theories. In particular, it provides a pathway to theories that may not emerge from a covariant Lagrangian formulation. This distinction highlights the role of Hamiltonian methods in theories beyond general relativity (GR). By focusing on the canonical structure rather than enforcing covariance at the Lagrangian level, this approach opens new possibilities for formulating gravitational theories that retain key features of GR while allowing for modifications that may not be accessible through traditional Lagrangian constructions.} The price to pay for this explicit covariance is losing the direct contact with holonomy corrections, and whether these last works are actual effective models of LQG is still an open question. Nevertheless, the modifications have a solid foundation as they ensure that the semiclassical theory has a univocal geometric description.

We want to describe the complete spacetime picture of a collapsing spherical star surrounded by vacuum. Following Ref.~\cite{Alonso-Bardaji:2023qgu}, we will minimally couple the dustlike matter to the effective Hamiltonian in such a way that both the vacuum and matter solutions are free of curvature singularities. Then, we will study the conditions for the matching (with no thin shells) of both geometries across a hypersurface comoving with the flow of matter. The resulting effective theory is parametrized by the polymerization constant of the holonomy corrections. It describes a geodesically complete spacetime, where the singularity is replaced with a boundary of finite curvature in the interior of the black hole. 

Let us point out that such a replacement has already been studied as a candidate to resolve the singularities in theories beyond general relativity \cite{Carballo_Rubio_2020}, and these geometries resemble some solutions in conformal gravity, where singularities are known to be absent outside the Einstein frame \cite{Naelikar_1977,Bambi_2017}. It is interesting to point out that similar solutions (as opposed to the usual bounce to white-hole spacetimes) have been suggested in LQG for the symmetry-reduced Schwarzschild interior \cite{B_hmer_2007,Alesci_2020}, {where the black-hole interior evolves into a de Sitter universe. Still, in this effective model, the area of the spheres remains finite on the novel boundary, which resembles more the previous studies in Refs.~\cite{Han:2022rsx,Han:2020uhb}. In these cases, the core of the black hole is an asymptotically (spacelike) Nariai-like geometry, that is, a de Sitter null infinity foliated by spheres of minimum area.}

The organization of the article is as follows. In Sec.~\ref{sec.ham} we introduce the effective Hamiltonian. In Sec.~\ref{sec.star}, we study the vacuum (Sec.~\ref{subsec.vac}) and matter (Sec.~\ref{subsec.dust}) solutions associated with this Hamiltonian, and, in Sec.~\ref{sec.matching}, we prove that both solutions can be matched across a timelike hypersurface, giving rise to the spacetime depicted in Fig.~\ref{fig.complete}. Sections \ref{sec.dust} and \ref{sec.vac} are devoted to a more profound analysis of the interior (effective dust) and exterior (vacuum) regions of the whole spacetime, respectively, with a particular emphasis on the global structure and singularity resolution. Finally, in Sec.~\ref{sec.evaporation}, we study the formation of black-hole remnants, and we end with a brief summary in Sec.~\ref{sec.conclusions}.

\section{The Effective Theory}\label{sec.ham}

Holonomy corrections have been shown to correctly reproduce the full quantum dynamics of loop quantum cosmology, replacing the big bang singularity with a quantum bounce at early times \cite{Ashtekar_2006_1,Ashtekar_2006, Rovelli_2014, Diener_2014}. In brief, these corrections replace the curvature or connection components with their imaginary exponential form. This leads to changes of the form $k\to\sin(\lambda k)/\lambda$, where the polymerization parameter $\lambda$ might be constant (scale independent) or a function of phase-space variables (scale dependent). However, in spherical symmetry, such simple models conflict with the explicit covariance of GR. While it is unclear whether they might descend from some symmetry-reduced covariant theory, they cannot describe a single geometry. In other words, {each} gauge solution to the {same} Hamiltonian equations produces a {different} metric (and not just a different chart of the same geometry), where singularities might or might not be present, with a different number of horizons, etc. To overcome this situation the modifications must be more profound. The search for a covariant effective {Hamiltonian for} spherical loop quantum gravity emerged as an ambitious project \cite{Bojowald_2015,Tibrewala_2014,Arruga_2020}, and recently produced the first positive results \cite{Alonso-Bardaji:2021yls,Alonso-Bardaji:2022ear,Alonso-Bardaji:2023niu}. These were soon generalized to include further modifications \cite{Alonso-Bardaji:2023vtl,Bojowald:2023xat} and even matter with dynamical degrees of freedom \cite{Alonso-Bardaji:2023qgu}. In the present work we take the baton of this last work and suggest a modified theory with scale-independent holonomy corrections that admits minimally coupled matter, and describes a nonsingular gravitational collapse. At the core of the calculation lies that
\begin{enumerate}[label=\textit{(\roman*)}]
    \item  Each gauge solution provides the same metric (so that each resulting chart is related to the others by a suitable coordinate transformation).
\end{enumerate}
In spherical symmetry, the spacetime is a warped product between a two-dimensional Lorentzian manifold and the two-sphere. We will require that
\begin{enumerate}[label=\textit{(\roman*)}]
\setcounter{enumi}{1}
     \item The time and radial functions define the Lagrange multipliers (lapse and shift) in the same way as in GR. 
\end{enumerate}
More precisely, the Lagrange multipliers are exactly the lapse and the shift components of the metric. Then, the metric (if any) on the Lorentzian part is completely determined by the algebra between constraints. On the other hand, the information about the angular component multiplying the metric of the unit sphere is lost due to the spherical symmetry reduction (any scalar function can play this role). Sticking to a minimal-modification prescription, we will further assume that 
\begin{enumerate}[label=\textit{(\roman*)}]
\setcounter{enumi}{2}
    \item The corrections preserve the area of the orbits of the rotation group.
\end{enumerate}
That is, we enforce the same area-radius function as in GR. We will say that a metric is associated with the Hamiltonian provided the above three conditions are satisfied, and this metric (if it exists) is unique. Let us specify the additional condition for its existence. 

In the canonical formalism, the total Hamiltonian $H_T$ is a linear combination of the (three) spatial diffeomorphism constraints and the Hamiltonian constraint. Choosing coordinates adapted to the spherical symmetry, the two angular components of the diffeomorphism constraint are identically vanishing, and the total Hamiltonian $H_T:=H[N]+D[N^x]$ follows
\begin{subequations}\label{eq.hdageneral}
  \begin{align}
 \big\{D[f_1],D[f_2]\big\}&=D\big[f_1f_2'-f_1'f_2\big],\\
 \big\{D[f_1],{H}[f_2]\big\}&={H}\big[f_1f_2'\big],\\
  \big\{{H}[f_1],{H}[f_2]\big\}&=D\left[
    F(f_1f_2'-f_1'f_2)\right],
  \end{align}
\end{subequations}
with $D[f]:=\int f\diff dx$ and $H[f]:=\int f\ham dx$. This must have an associated metric of the form \cite{Teitelboim73, Alonso-Bardaji:2023vtl},
\begin{align}\label{eq.metricgoodgen}
    ds^2=\varsigma N^2dt^2 +\frac{1}{|F|}\big(dx+N^xdt\big)^2+r^2 d\sigma^2 ,
\end{align}
with $\varsigma:=-\mathrm{sgn}(F)$ the spacetime signature, $d\sigma^2$ the metric on the unit sphere, {and $F$ and $r$ phase-space functions}. Of course, the metric components must transform in a very specific way \cite{Pons_1997}:
\begin{subequations}\label{eq.covarianceall}
\begin{align}
\label{gaugelapse}&\partial_t(\xi^t N) +\xi^x\partial_x N -N N^x\partial_x\xi^t \overset{os}{=}\partial_t{\epsilon} +\epsilon^x\partial_x N -N^x\partial_x\epsilon\\
\label{gaugeshift}&\partial_t{(\xi^t{N^x}+\xi^x)} +{\xi^x}\partial_x N^x-N^x\partial_x\xi^x-\left((N^x)^2 +N^2F\right){\partial_x{\xi^t}}\overset{os}{=}\partial_t{\epsilon}^x +\epsilon^x\partial_x N^x -N^x\partial_x\epsilon^x +{F}(\epsilon\partial_x N -N\partial_x\epsilon)\\
     & \xi^t\partial_t({{1}/{F}})+\xi^x\partial_x ({{1}/{F}}) +({{2/}{F}})\left({N^x}\partial_x{\xi^t}+\partial_x{\xi^x}\right)\overset{os}{=}\{(1/F), H[\epsilon]+D[\epsilon^x]\},\label{eq.covariance}\\
     & \xi^t\partial_t(r^2)+\xi^x\partial_x (r^2) \overset{os}{=}\{r^2, H[\epsilon]+D[\epsilon^x]\},\label{eq.covariancescalar}
\end{align}
\end{subequations}
where ``$os$'' denotes that the equalities must hold on shell. This means that infinitesimal coordinate changes [described by the Lie derivative along the vector field $\xi^\mu=(\xi^t,\xi^x,0,0)$ on the left hand side] coincide with the suitable gauge transformations of the spatial metric components as phase-space functions (those where the coefficients of $H$ and $D$ are the normal and tangential components of $\xi$, respectively, in the coordinate basis adapted to the foliation, i.e., $ \epsilon:=\xi^t N$ and $\epsilon^x:=\xi^tN^x+\xi^x$). The same applies for the Lagrange multipliers that we identify with the lapse and shift (this is not the only possibility; see, for instance, Appendix~\ref{app.extra} or Ref.~\cite{Alonso-Bardaji:2023qgu}). The above relations are then the additional necessary condition for the existence of the metric \eqref{eq.metricgoodgen} as an associated object to the (covariant) Hamiltonian.

Of course, the GR Hamiltonian satisfies all these conditions as it descends from an explicitly covariant Lagrangian. However, this is not obvious in modified Hamiltonian models. {In fact, the role of diffeomorphism symmetry in canonical theories beyond GR is a topic under current investigation, particularly in the context of LQG, where the canonical and covariant approaches provide complementary perspectives on quantum gravity \cite{Han:2009bb}. It is yet not clear how to reconcile the former with a spacetime formulation. In this work, we continue working on alternative approaches to generally covariant theories in the Hamiltonian formalism.} 

In Refs.~\cite{Alonso-Bardaji:2021yls,Alonso-Bardaji:2022ear}, we presented the first model satisfying all the above conditions and proved that the Schwarzschild singularity is resolved by scale-independent covariant holonomy corrections. Nonetheless, this result turned out to be weak under the addition of minimally coupled matter with local degrees of freedom \cite{Bojowald:2023djr}. As proven in Ref.~\cite{Alonso-Bardaji:2023qgu}, this is not the end of the story. In fact, there is an infinite family of ``conformal'' Hamiltonians that admit minimally coupled (effective) dust in such a way that spacetime singularities are removed. Out of this whole family, a particular case is singled out by basic physical assumptions, as we show in the following. Further motivation and relation with previous models can be found in Appendix~\ref{app.extra}. In the next subsection we present the Hamiltonian and associated metric of the effective theory.

\subsection{Effective covariant Hamiltonian}

The symplectic structure of the phase space is given by the two canonical pairs
\begin{align}\label{eq.vars}
    \{\krad(x_a),\erad(x_b)\}=\delta(x_a-x_b)=\{\kang(x_a),\ephi(x_b)\},
\end{align}
and the effective Hamiltonian will be a linear combination of the constraints
\begin{subequations}\label{polham}
\begin{align}
\label{eq.polhamdiff}\diff_g &= -\erad'\krad+\ephi\kang' ,\\
    \label{eq.polhamham}\ham_g &=  \frac{\Omega(\erad,m)}{\sqrt{1+\lambda^2}}\Bigg[-\frac{{\ephi}}{2\sqrt{{\erad}}}\left(1+\frac{\sin^2{{(\lambda {\kang})}}}{{{\lambda^2}}}\right) +\left(\frac{({\erad}')^2}{8\sqrt{{\erad}}{\ephi}}+\frac{\sqrt{{E}^x}}{2}\left(\frac{{E}^x{}'}{{E}^\varphi}\right)'\right)\cos^2{(\lambda {\kang})} \nonumber\\
    &\qquad\qquad\qquad\;-\sqrt{{\erad}}{\krad}\frac{\sin{(2\lambda {\kang})}}{\lambda}\left(1+\left(\frac{\lambda {\erad}'}{2{\ephi}}\right)^{\!2}\right)\Bigg],
\end{align}
\end{subequations}
with $\lambda\in\mathbb{R}^+$ the polymerization parameter. The only difference with respect to the work in Refs.~\cite{Alonso-Bardaji:2021yls,Alonso-Bardaji:2022ear} lies on the global factor $\Omega=\Omega(\erad,m)$, which is a free function of the scalar $\erad$ and the {phase-space function},
\begin{align}
    \label{eq.defm}  m:=\frac{\sqrt{\erad}}{2}\left(1+\frac{\sin^2(\lambda\kang)}{\lambda^2}-\left(\frac{\erad'}{2\ephi}\right)^{2}\cos^2(\lambda\kang)\right).
\end{align}
{The relevance of this last function is that its (spatial) derivative, $m'$, is a linear combination of the vacuum constraints, and thus it is a conserved quantity on the constraint surface (in vacuum). In fact, it coincides with the Schwarzschild mass in the GR limit.} 

The above constraints satisfy the hypersurface deformation algebra \eqref{eq.hdageneral} with the everywhere non-negative structure function  
\begin{align}\label{eq.structurefk}
    F:=\Omega^2\frac{\cos^2(\lambda\kang)}{1+\lambda^2}\left[1+\left(\frac{\lambda \erad'}{2{\ephi}}\right)^{\!2}\right]\frac{\erad}{\ephi^2},
\end{align}
which is just $\Omega^2$ times the structure function in Refs.~\cite{Alonso-Bardaji:2021yls,Alonso-Bardaji:2022ear}, because the scalar $\Omega$ commutes with itself and anticommutes with the Hamiltonian constraint {off shell}. One can check that this follows the covariance condition \eqref{eq.covariance}. 

Then, the metric associated with the Hamiltonian \eqref{polham} is just \eqref{eq.metricgoodgen} with a constant signature $\varsigma=-1$, the function $F$ as defined above, and $r^2=\erad$ due to condition {(iii)} at the beginning of the section, which, of course, satisfies \eqref{eq.covariancescalar}. Using the function~\eqref{eq.defm}, we may rewrite 
\begin{align}\label{eq.structurefm}
    F= \Omega^2\left(1-\frac{2\lambdabar m}{\sqrt{\erad}}\right)\frac{\erad}{\ephi^2} ,
\end{align}
where the constant $\lambdabar:=\lambda^2/(1+\lambda^2)$ is bounded between zero and one because the polymerization parameter $\lambda$ is a nonvanishing real constant. This represents the strength of the quantum corrections on the geometry. The limits $\lambdabar\to0$ and $\lambdabar\to1$ correspond to classical GR and the maximum departure from Einstein's theory, respectively. We thus expect $\lambdabar\ll1$ in astrophysical scenarios. 

The resulting spacetime is Lorentzian, with metric
\begin{align}\label{eq.metricfinal}
    ds^2=-{N}^2dt^2+\Omega^{-2}\left(1-\frac{2\lambdabar m}{\sqrt{\erad}}\right)^{-1}\frac{\ephi^2}{\erad}\big(dx+N^xdt\big)^2+\erad d\sigma^2. 
\end{align}
To describe the dynamical collapse, we need to incorporate matter with local degrees of freedom. In these models, it is possible to follow the usual minimal coupling prescription, as shown in Refs.~\cite{Alonso-Bardaji:2023vtl,Alonso-Bardaji:2023qgu}. We introduce a third pair of canonical variables
\begin{align}
    \{\phi(x_a),\pphi(x_b)\}=\delta(x_a-x_b),
\end{align}
and add the matter constraints
\begin{subequations}\label{eq.matterham}
    \begin{align}
        \diff_m&:=\phi'\pphi,\\
        \ham_m&:=\pphi\sqrt{1 +F({\phi}')^2},
    \end{align}
\end{subequations}
to \eqref{eq.polhamdiff} and \eqref{eq.polhamham}, respectively. The total Hamiltonian reads
\begin{align}\label{eq.hampoltot}
    H_T=\int\big(N(\ham_g+\ham_m)+N^x(\diff_g+\diff_m)\big)dx,
\end{align}
with the constraints as given in \eqref{polham} and \eqref{eq.matterham}. By construction, it satisfies the algebra \eqref{eq.hdageneral} with the structure function \eqref{eq.structurefm} and the metric \eqref{eq.metricfinal}. As expected, on shell, $m'$ is a linear combination of the matter constraints only, that is,
\begin{align}\label{eq.mprimematter}
    m'\approx \sqrt{1+\lambda^2}\frac{\erad'}{2\ephi}\frac{\ham_m}{\Omega}-\frac{\partial m}{\partial\kang}\frac{\diff_m}{\ephi}=\pphi\left( \frac{\sqrt{1+\lambda^2}}{\Omega}\frac{\erad'}{2{\ephi}}\sqrt{1+F{(\phi')^2}} -\frac{\phi'}{\ephi}\frac{\partial m}{\partial\kang}\right),
\end{align}
where ``$\approx$'' denotes that the equality holds on the constraint surface. Then, the vacuum limit corresponds to either $\pphi=0$ or, equivalently, $m'=0$ on shell. {Most remarkably, even if $m$ is no longer a constant of motion [in fact, it is just a specific definition for the phase-space function \eqref{eq.defm}], the covariance conditions are satisfied for any (scalar) function $\Omega(\erad,m)$. We refer the interested reader to Appendix~\ref{app.extra} for further details.}

\section{The Collapsing Star}\label{sec.star}

In this section we study the solution to the Hamiltonian equations given by \eqref{eq.hampoltot}. We begin with the vacuum case, where there are only two pairs of canonical variables and no propagating degrees of freedom. In Sec.~\ref{subsec.vac}, we will show that the resulting family of geometries is completely determined by the two constant parameters $m$ and $\lambdabar$ for any choice of the global factor $\Omega$. We then turn to the matter solution in Sec.~\ref{subsec.dust}. In this case, we will choose the matter reference frame, where the dynamics gets greatly simplified, and the GR limit becomes evident. Last, we study the geometric matching of these two spacetimes in Sec.~\ref{sec.matching}, and obtain the junction conditions so that it can be performed with no thin shells. These requirements, plus the asymptotic flatness of the vacuum solution, fix the function $\Omega$ up to a constant, which can be chosen so that the spacetime is free of singularities. 

In Sec.~\ref{sec.dust} and Sec.~\ref{sec.vac}, we will study in further detail the nonsingular matter and vacuum geometries after fixing the global factor.

\subsection{Vacuum reduction}\label{subsec.vac}

The vacuum solution is very similar to the one described in Refs.~\cite{Alonso-Bardaji:2021yls,Alonso-Bardaji:2022ear}, where the Schwarzschild singularity is replaced with a totally geodesic spacelike surface of finite curvature (foliated by spheres of minimum area) in the trapped region of the spacetime. In fact, the only difference with the present model is a conformal factor in the Lorentzian sector. We prove this in Appendix~\ref{app.extra}. Here, we only show it by explicitly solving the Hamiltonian equations.

We choose the radial-polar gauge, i.e., $\erad=x^2$ and $\kang=0$, because the system reduces to the same equations as in GR except for the new global factor $\Omega$. From the diffeomorphism-constraint equation, we see that $\krad=0$, and $\dot{\erad}:=\{\erad,H_T\}=0$ sets $N^x=0$. The mass function \eqref{eq.defm} can be solved for $\ephi=x/\sqrt{1-2m/x}$, which then satisfies the Hamiltonian constraint. The conservation of this condition, $\dot{\ephi}:=\{\ephi,H_T\}=0$, provides
\begin{align}\label{eq.lapse}
    N=\frac{c}{\Omega}\sqrt{1-\frac{2m}{x}} ,
\end{align}
with the constant $c$ a trivial redefinition of time, and we set it to one. With a small abuse of notation, we relabel $x$ as $r$, and write the metric in the chart $\{t,r\}$,
\begin{align}\label{eq.metricdiagvac}
    ds^2=\Omega(r,m)^{-2}\Bigg[-\left(1-\frac{2 m}{r}\right)dt^2+\left(1-\frac{2\lambdabar m}{r}\right)^{-1}\left(1-\frac{2 m}{r}\right)^{-1}dr^2\Bigg]+r^2d\sigma^2.
\end{align}
To find the suitable fall-off conditions for asymptotic flatness, we need $\lim_{r\to\infty}\Omega=1+\mathcal{O}(1/r)$. The limit $\lambdabar\to0$ yields the Schwarzschild metric plus the factor $\Omega^{-2}$. But in contrast to the effective model, we have the fundamental (Einstein's) field equations, and these impose $\lim_{\lambdabar\to0}\Omega=1$ to recover GR. As anticipated above, this metric is just the one presented in Refs.~\cite{Alonso-Bardaji:2021yls,Alonso-Bardaji:2022ear} with the additional conformal factor in the Lorentzian sector. If $\Omega$ has no roots or poles bigger than $2m$ (which would imply drastic quantum corrections in classical regions), the chart is valid for $2m<r$; otherwise, until the biggest root or pole of $\Omega$.

\subsection{The matter solution}\label{subsec.dust}

As already specified above, the mass function is no longer a constant of motion when adding matter, but its derivative \eqref{eq.mprimematter} gets greatly simplified when working in the matter reference frame $\phi=t$. If we turn to the equation of motion $\dot{\phi}:=\{\phi,H_{T}\}=N^x\phi'+N\sqrt{1+F(\phi')^2}=1$, we can read $N=1$ in this gauge. Then, the conjugate variable $\pphi$ is the energy, and we can divide it by the determinant of the spatial metric to compute the energy density,
\begin{align}\label{eq.energydensity}
    \rho:= \frac{\sqrt{F}}{\erad}\pphi\approx 2\sqrt{1-\lambdabar}\left({1-\frac{2\lambdabar m}{\sqrt{\erad}}}\right)^{\!1/2}\!\frac{\Omega^2 m'}{\sqrt{\erad}\erad'} \,,
\end{align}
which depends only on the scalar functions $\erad$, $m$, and their first-order spatial derivatives.

After this partial gauge fixing, there are still four equations of motion ($\dot{\erad},\dot{\ephi},\dot{\krad},\dot{\kang}$) and the diffeomorphism constraint. To ease the resolution, we solve $\diff=0$ for $\krad=\ephi\kang'/\erad'$ (assuming a nonidentically vanishing $\erad'$), and we perform the change of variables $(\erad,\ephi,\kang)\to(r,m,\kappa)$, with $r:=\sqrt{\erad}$, $m$ as defined in \eqref{eq.defm}, and
\begin{align}\label{eq.kappadef}
    \kappa&:=\left(\frac{\erad'}{2\ephi}\right)^{2}\frac{1+\lambda^2\sin^2(\lambda\kang)}{1+\lambda^2}-\frac{\cos^2(\lambda\kang)}{1+\lambda^2}=\left(\frac{rr'}{\ephi}\right)^2-1+\frac{2\lambdabar m}{r}.
\end{align}
We want to remark that this function $\kappa$ differs from that in Ref.~\cite{Alonso-Bardaji:2023qgu} by the factor $2\lambdabar m/r$. Of course, the dynamics is not altered by this choice, but it is more convenient in order to study the matching, as we will see in the next section. 
In terms of the new variables, the three remaining evolution equations get greatly simplified,
\begin{subequations}\label{eq.dotn}
\begin{align}
        \label{eq.rdotn}\dot{r}&=\shift r' -\epsilon\, \Omega \sqrt{1-\frac{2\lambdabar m}{r}}\sqrt{\kappa+(1-\lambdabar)\frac{2m}{r}},\\
        \label{eq.mdotn}\dot{m}&=\shift m' ,\\
        \label{eq.kdotn}\dot{\kappa}&=\shift\kappa' -\epsilon\,\Omega\sqrt{1-\frac{2\lambdabar m}{r}}\sqrt{\kappa+(1-\lambdabar)\frac{2m}{r}}\Bigg(\left(1+\kappa-\frac{2\lambdabar m}{r}\right)\frac{\partial\log(\Omega^2)}{\partial r}-\frac{2\lambdabar m}{r^2}\Bigg),
\end{align}
\end{subequations}
{with $\Omega=\Omega(r,m)$ and} where $\epsilon:=-\mathrm{sgn}(\lambda\sin(2\lambda\kang))$ is dynamical. It changes whenever $\sin(2\lambda\kang)$ vanishes (because $\dot{\kang}\neq0$) there. These points correspond to $r=2\lambdabar m$ (roots of the cosine) and $r=2m(\lambdabar-1)/\kappa$ (roots of the sine, which are only attainable for negative $\kappa$). This can be seen by inserting \eqref{eq.kappadef} in \eqref{eq.defm}, and then setting $\cos(\lambda\kang)=0$ and $\sin(\lambda\kang)=0$, respectively. 

We can fix one more gauge freedom, and this will lead us to a chart plus two first-order evolution equations. Our choice is to set a time-independent mass profile, $m=m(x)$, which sets $\shift=0$ through \eqref{eq.mdotn}. Note that in GR this would also imply a time-independent $\kappa$, but not in the present scenario, where both $\Omega$ and the constant $\lambdabar$ ``spoil'' this result. Nevertheless, we will find that the matching conditions impose that $\kappa$ is constant over the matching hypersurface, fixing in that way the function $\Omega$.

After relabeling $t$ as the function $\tau$ on the manifold, the metric \eqref{eq.metricfinal} in the chart $\{\tau,x\}$ reads
\begin{align}\label{eq.metricdiag2+}
    ds^2=&\,-{d\tau}^{2}+ \Omega^{-2}\left(1-\frac{2\lambdabar m}{r}\right)^{-1}\frac{\big(r'\big)^2}{1+\kappa-2\lambdabar m/r}\,{dx}^{2}+r^{2}{d\sigma}^{2},
\end{align}
with $\Omega=\Omega(r,m)$, $m=m(x)$, $r=r(\tau,x)$, and $\kappa=\kappa(\tau,x)$. The evolution equations \eqref{eq.dotn} reduce to
\begin{subequations}\label{eq.dotn2}
\begin{align}
        \label{eq.rdotn2}\dot{r}&=-\epsilon\, \Omega \sqrt{1-\frac{2\lambdabar m}{r}}\sqrt{\kappa+(1-\lambdabar)\frac{2m}{r}},\\
        \label{eq.mdotn2}\dot{m}&=0,\\
        \label{eq.kdotn2}\dot{\kappa}&=\mathcal{A}\,\dot{r},
\end{align}
\end{subequations}
with 
\begin{align}\label{eq.anomalykappa}
      \mathcal{A}:=\left(1+\kappa-\frac{2\lambdabar m}{r}\right)\frac{\partial\log(\Omega^2)}{\partial r}-\frac{2\lambdabar m}{r^2}.
\end{align}
From the first one, we see that $\epsilon=+1$ ($\epsilon=-1$) corresponds to the ``contracting'' (``expanding'') branch. 

Now, we want to understand whether the divergences of the above line element describe curvature singularities. The surfaces where the chart breaks down are $r'=0$, $r=2\lambdabar m/(1+\kappa)$, $r=2\lambdabar m$, and the roots of $\Omega$. The first two coincide by definition, and thus  $r=2\lambda^2m/(\lambda^2+\sin^2(\lambda\kang))\in[2\lambdabar m,2m]$. This suggests that they are equivalent to the usual shell-crossing singularities in LTB models. Even if curvature scalars diverge there, as long as $r'>0$, these weak singularities can be avoided. Further work to study specific mass profiles and shell-crossing singularities is expected in the near future. The roots of $\Omega$ do not involve curvature divergences. Still, its derivatives might spoil this result, and we would need to study the behavior for each specific function. Last, the surfaces $r=2\lambdabar m$ are true physical singularities, and the only way to remove them is by forcing the yet free function $\Omega$ to vanish at least as $\sqrt{r-2\lambdabar m}$ when approaching this surface, as explained in Appendix~\ref{app.extra}. In the next section we will see that this decay is compatible with the matching conditions on the surface of the star.

\subsection{Geometric matching}\label{sec.matching}

{In the previous subsections, we found the vacuum \eqref{eq.metricdiagvac} and nonvacuum \eqref{eq.metricdiag2+}--\eqref{eq.dotn2} solutions to the effective Hamiltonian \eqref{eq.hampoltot}. In this subsection, we extract the necessary conditions for these two families of spacetimes to describe a physically meaningful spherical star, that is, a configuration of matter surrounded by vacuum. In other words, we impose and solve the Israel junction conditions \cite{Israel_1966}. We will see that the yet free function $\Omega$ plays a crucial role, as not every choice is compatible with the junction conditions. However, the allowed forms of $\Omega$ [see Eq.~\eqref{eq.defomega}] lead to singularity-free spacetimes, as we will prove in the following sections. The rest of this subsection presents the detailed computation of the junction conditions, which are summarized in \eqref{eq.landm} and \eqref{eq.kappaandomega}. The geometry of the matching hypersurface (i.e., the surface of the star) is described in \eqref{eq.matchinghypersurface}.}

With both the matter ($\mathcal{M}^+,g^+$) and vacuum ($\mathcal{M}^-,g^-$) spacetimes at hand, we want to identify them with some ``internal'' and ``external'' solutions, so that the whole spacetime ($\mathcal{M},g$) is defined as the disjoint union of $\mathcal{M}^+$ and $\mathcal{M}^-$. For that purpose, we need two diffeomorphic timelike hypersurfaces $\Sigma^\pm$ (one for each solution), with the same image $\Sigma$ on $\mathcal{M}$. This resulting spacetime has a Lorentzian geometry provided the first fundamental forms on $\Sigma^\pm$ (the induced metrics) coincide~\cite{Israel_1966,Mars_1993}. In addition, there are no shell distributions on the curvature tensors if the second fundamental forms on $\Sigma^\pm$ (the extrinsic curvatures) are the same. 

We assume that the matching preserves the spherical symmetry \cite{Vera_2002}, so that the metrics read $g^\pm=g_{(2)}^\pm+r_\pm^2d\sigma^2$, and we omit the angular coordinates in the following. Then, the matching reduces to a two-dimensional problem, and the boundaries are characterized as $\Sigma^{\pm}:=\{x^0_\pm=T_s^\pm(l),x^1_\pm=R_s^\pm(l)\}$, with the (timelike) vectors 
\begin{subequations}
\begin{align}
     e^\mu_\pm\partial_\mu=\frac{dT_s^\pm}{dl}\partial_{x^0_\pm}+\frac{dR_s^\pm}{dl}\partial_{x^1_\pm},
\end{align}
as functions of $l$, being tangent to them. The (spacelike) unit normals are unique up to the overall sign $\varepsilon^\pm$,
\begin{align}\label{eq.normaltosigma}
    n_\mu^\pm dx_\pm^\mu=\varepsilon^\pm\sqrt{\frac{\mathrm{det}\big(g^\pm_{(2)}\big)}{e^\pm_\mu e_\pm^\mu}}\left(
    -\frac{dR_s^\pm}{dl}dx^0_\pm+\frac{dT_s^\pm}{dl}dx^1_\pm
    \right),
\end{align}
\end{subequations}
and both sets of vectors $\{n_\pm^\mu\partial_\mu,e_\pm^\mu\partial_\mu\}$, plus the two vectors on the spheres, provide complete bases of the tangent space of $\mathcal{M}$ on $\Sigma$. We now now need to identify the images of $\Sigma^\pm$ on $\mathcal{M}$. 

From the equality of the first and second fundamental forms,
\begin{subequations}
\begin{align}
    h^\pm     &:= e_\mu^\pm e^\mu_\pm dl^2+r_\pm^2 d\sigma^2,\\
    k^\pm     &:=-n^\pm_\mu e_\pm^\nu\nabla^\pm_\nu e_\pm^\mu dl^2 + r_\pm n^\pm_\mu \nabla_\pm^\mu r_\pm d\sigma^2,
\end{align}
\end{subequations}
we can read the four independent matching conditions 
\begin{subequations}\label{eq.matching}
\begin{align}
    e^+_\mu e_+^\mu&\overset{\Sigma}{=}e_\mu^- e_-^\mu,\label{eq.matching1}\\
    r_+&\overset{\Sigma}{=}r_-,\label{eq.matching2}\\
    n^+_\mu e_+^\nu\nabla^+_\nu e_+^\mu&\overset{\Sigma}{=}n^-_\mu e_-^\nu\nabla^-_\nu e_-^\mu,\label{eq.matching3}\\
    n^+_\mu \nabla_+^\mu  r_+&\overset{\Sigma}{=}n^-_\mu \nabla_-^\mu  r_- .\label{eq.matching4}
\end{align}
\end{subequations}
In particular, \eqref{eq.matching2} ensures that the area-radius function takes the same value from both sides at any point of $\Sigma$. Its derivative along the tangent vectors $e_\pm^\mu\partial_\mu$ plus condition \eqref{eq.matching4} then imply the continuity of the whole gradient $\nabla r$. This can be expressed as \cite{Fayos_1996,Mimoso_2010},
\begin{align}\label{eq.matching4b}
    M_H^+\overset{\Sigma}{=}M_H^-,
\end{align}
where $M_H$ is the Hawking mass (or, equivalently, the Misner-Sharp mass in spherical symmetry). It is remarkable that it has the same functional form in both spacetimes,
\begin{align}\label{eq.hawkingmass}
   M_H:=\frac{r}{2}\big(1-(\nabla_\mu r) (\nabla^\mu r)\big)=\frac{r}{2}\left(1-\left(1-\frac{2m}{r}\right)\left(1-\frac{2\lambdabar m}{r}\right)\Omega(r,m)^2\right)  .
\end{align}
Note that in GR, this immediately implies the equality of the parameter $m$ from both sides. In this case, however, the identification is not straightforward due to the additional factor $\lambdabar$ and the dependence on $\Omega$. 

We leave the general case for future work, and, here, we check whether the matching can be performed across the hypersurface comoving with the flow of matter (because this is the ``physically meaningful'' scenario for a collapsing star). In the chart \eqref{eq.metricdiag2+}, $\Sigma^+$ is thus defined by a constant $x$ and $\tau= l$, i.e.,
\begin{align}
  e^\mu_+\partial_\mu =\partial_{\tau} &&\mathrm{and}&&    n_\mu^+dx^\mu_+= (g_{xx}^+)^{1/2}\,dx,
\end{align}
where we have set $\varepsilon^+=1$, yielding $h^+(\partial_l,\partial_l)=-1$ and $k^+(\partial_l,\partial_l)=0$. Besides, the function $m^+=m^+(x)$ is constant over the matching hypersurface, as one can read from \eqref{eq.mdotn2}.
Since $m^-$ and $\lambdabar^\pm$ are also constants, the only solution to \eqref{eq.matching4b} that is compatible with a continuous limit to GR ($\lambdabar^\pm\to0$ and $\lim_{\lambdabar^\pm\to0}\Omega=1$) is 
\begin{subequations}\label{eq.landm}
\begin{align}
   m^+\overset{\Sigma}{=}m^-   .
\end{align}
That is, a constant and equal mass parameter on the matching hypersurface (just as in GR). As a result, $\Omega(r,m)$ is continuous across the matching hypersurface, and we must have
\begin{align}
    \lambdabar^+\overset{\Sigma}{=}\lambdabar^-.
\end{align}
\end{subequations}
As we show next, $\Sigma^-$ is already completely determined in $(\mathcal{M}^-,g^-)$. Let us work with the chart \eqref{eq.metricdiagvac}. From \eqref{eq.matching4},
\begin{subequations}
    \begin{align}\label{eq.dtdl}
        \frac{dT_s^-}{dl}&=\varepsilon^-\Omega\sqrt{1+\kappa-\frac{2\lambdabar m}{r}}\left(1-\frac{2m}{r}\right)^{-1}\bigg|_{\Sigma},
    \end{align}
and taking the derivative of \eqref{eq.matching2} with respect to $l$,
\begin{align}
     \frac{dR_s^-}{dl}&=-\epsilon\Omega\sqrt{1-\frac{2\lambdabar m}{r}}\sqrt{\kappa+(1-\lambdabar)\frac{2m}{r}}\bigg|_{\Sigma},
\end{align}
\end{subequations}
where we used \eqref{eq.rdotn2}. The sign $\varepsilon^-=\pm1$ stands for the relative orientation of the time coordinates in $\mathcal{M}^+$ and $\mathcal{M}^-$. It is straightforward to check that this satisfies \eqref{eq.matching1}, and the only remaining matching condition is \eqref{eq.matching3}. Using all the above, it reduces to 
\begin{align}
e_+^\mu\partial_\mu\kappa\overset{\Sigma}{=}0.    
\end{align}
Since $e^\mu_+\partial_\mu =\partial_{\tau}$, this is equivalent to $\dot{\kappa}=0$ on the matching hypersurface. From \eqref{eq.kdotn2}, we read $\mathcal{A}|_{\Sigma}=0$, because we are interested in the dynamical collapse, and thus $\dot{r}$ cannot vanish everywhere on $\Sigma$. Therefore, the function $\mathcal{A}$, as given in \eqref{eq.anomalykappa}, must vanish on the matching surface. We can solve this for
\begin{align}
    \kappa\overset{\Sigma}{=} \frac{2\lambdabar m}{r}\left(1+\frac{1}{rf}\right)-1,
\end{align}
with $f=\partial\log(\Omega^2)/\partial r$. 
Clearly, the derivatives of $\mathcal{A}$ along $\Sigma$ are also zero, and from
\begin{align}
    e^\mu\partial_\mu\mathcal{A} =\left[\left(\mathcal{A}+\frac{2\lambdabar m}{r^2}\right)f + \left(1+\kappa-\frac{2\lambdabar m}{r}\right)\frac{\partial f}{\partial r}+\frac{4\lambdabar m}{r^3}\right]\dot{r}\overset{\Sigma}{=}\frac{2\lambdabar m}{r^2}\left(f+\frac{\partial f}{\partial r}+\frac{2}{r}\right)\dot{r}\overset{\Sigma}{=}0,
\end{align}
we obtain $f|_\Sigma=2\lambdabar m/(r(b r-2\lambdabar m))$, with an integration constant $b$. Substituting above,
\begin{subequations}\label{eq.kappaandomega}
    \begin{align}\label{eq.solkappa}
        \kappa\overset{\Sigma}{=}b-1,
    \end{align}
so that $b\geq1$ from the definition \eqref{eq.kappadef}. 
Finally, direct integration of $f=\partial\log(\Omega^2)/\partial r$ yields
\begin{align}
     \Omega\overset{\Sigma}{=}a\sqrt{b-\frac{2\lambdabar m}{r}} ,
\end{align}
\end{subequations}
with constant $a\neq0$. 

Let us make a brief remark. In the GR limit, $\lambdabar=0$ and $\Omega=1$, the function $\kappa$ is a function of the radial coordinate $x$ everywhere on $\mathcal{M}^+$, so that it is constant on $\Sigma^+$, and it can be absorbed in a redefinition of the timelike Killing vector field of the vacuum spacetime ($\mathcal{M}^-,g^-$). This can be seen in \eqref{eq.dtdl}, where the term $\sqrt{1+\kappa(x)}$ (when $\lambdabar=0$) is a trivial rescaling of $dT_s^-/dl$. In the present case, the matching conditions also demand that $\kappa$ remains constant on the matching hypersurface, and we can promote $b$ in \eqref{eq.solkappa} to a function of $x$ to recover the GR result. 

But in this effective model, we have the additional function $\Omega$ that can be used to determine the constants $a$ and $b$. We may do so by enforcing the desirable physical properties of the spacetime solution. First, the asymptotic decay of $\Omega$ for the spacetime to be asymptotically flat fixes $b=1/a^2$. Second, the absence of singularities in the matter solution (see the end of the previous subsection) demands $b=1$, because $\Omega$ must vanish at least as fast as $\sqrt{r-2\lambdabar m}$ when $r\to2\lambdabar m$ (and this holds for any $x$). As a result, $\kappa|_\Sigma=0$. 

In summary, conditions \eqref{eq.landm} and \eqref{eq.kappaandomega} are enough for the existence of the matching between the matter and vacuum spacetimes along a hypersurface comoving with the flow of matter. These ensure that the normal projections of the Einstein tensor coincide from both sides~\cite{Mars_1993}. In particular, $n_+^\mu n^+_\nu (G_{\mu}{}^\nu)^+=n_-^\mu n^-_\nu (G_{\mu}{}^{\nu})^-$ and $e_+^\mu n^+_\nu (G_{\mu}{}^\nu)^+=e_-^\mu n^-_\nu (G_{\mu}{}^{\nu})^-$ on $\Sigma$. Dropping the ``$\pm$'' indices, the first and second fundamental forms on the timelike matching hypersurface~$\Sigma$ read
\begin{subequations}\label{eq.matchinghypersurface}
    \begin{align}
         h    &= -d\tau^2 +r^2 d\sigma^2,\\
     \label{eq.curv+}    k  &= r\left(1-\frac{2\lambdabar m}{r}\right)^{3/2}d\sigma^2.
    \end{align}
\end{subequations}
Still, there are some allowed (finite) discontinuities on the curvature, which are characterized by the difference of the projections of the corresponding Einstein tensors onto the matching hypersurface. In spherical symmetry there may be such two independent quantities:  $e_+^\mu e^+_\nu (G_{\mu}{}^\nu)^+-e_-^\mu e^-_\nu (G_{\mu}{}^{\nu})^-$ and $(G_\theta{}^\theta)^+-(G_\theta{}^\theta)^-$, with $\theta$ any of the coordinates on the sphere. In the GR limit $\lambdabar\to0$, the latter vanishes and the former is just $2m'/(r^2r')$.

In the following, we will describe in more detail the geometry of the spacetime after fixing the global function 
\begin{align}\label{eq.defomega}
    \Omega:=\sqrt{1-\frac{2\lambdabar m}{r}},
\end{align}
which, as we will show, ensures that the spacetime is geodesically complete. It also enforces a vanishing $\kappa$ on the surface of the star.

\section{The interior of the star}\label{sec.dust}

By considering \eqref{eq.defomega}, there are no curvature singularities as $r\to2\lambdabar m$. In fact, this represents a boundary of the trapped region of spacetime. The metric \eqref{eq.metricdiag2+} reads
\begin{align}\label{eq.metricdiag2}
    ds^2=&\,-{d\tau}^{2}+ \left(1-\frac{2\lambdabar m}{r}\right)^{\!\!-2}\!\!\frac{\big(r'\big)^2}{1+\kappa-2\lambdabar m/r}\,{dx}^{2}+r^{2}{d\sigma}^{2},
\end{align}
with $m=m(x)$, $r=r(\tau,x)$, and $\kappa=\kappa(\tau,x)$. They must follow the evolution equations, 
\begin{subequations}\label{eq.dot2}
\begin{align}
      \label{eq.rdot2}  \dot{r}&= -\epsilon\,\left(1-\frac{2\lambdabar m}{r}\right)\sqrt{\kappa+(1-\lambdabar)\frac{2m}{r}},\\
       \label{eq.kdot2}\dot{\kappa}&= -\epsilon\,\kappa\frac{2\lambdabar m}{r^2}\sqrt{\kappa+(1-\lambdabar)\frac{2m}{r}}.
\end{align}
\end{subequations}
If we focus on the evolution of the area-radius function \eqref{eq.rdot2}, we find that all time derivatives of $r$ vanish at the surface $r=2\lambdabar m$, and thus they cannot be attained in a finite amount of dust proper time. More precisely, the infimum $r=2\lambdabar m$ corresponds to the limit $\tau\to \epsilon\times\infty$. 
Note that the ``marginally bound'' case $\kappa=0$ is a valid solution for the dynamics, and it makes \eqref{eq.rdot2} directly integrable,
\begin{align}\label{eq.solparticular}
    \tau_0(x)-\tau&=\frac{2\epsilon}{\sqrt{1-\lambdabar}}\Bigg[\sqrt{\frac{r(\tau,x)}{2m(x)}}\left(\frac{1}{3}r(\tau,x)+2\lambdabar m(x)\right)
   -\lambdabar^{3/2} m(x) \log\left(\frac{\sqrt{r(\tau,x)}+\sqrt{2\lambdabar m(x)}}{\sqrt{r(\tau,x)}-\sqrt{2\lambdabar m(x)}}\right)\Bigg] ,
\end{align}
showing that in the collapsing branch ($\epsilon=1$) it takes an infinite amount of proper time to get to the infimum $r=2\lambdabar m$. 

For completeness, we also need to write down the energy density \eqref{eq.energydensity}, 
\begin{align}\label{eq.energydensitygauge}
    \rho=\sqrt{1-\lambdabar}\left({1-\frac{2\lambdabar m}{r}}\right)^{3/2}\frac{ m'}{r^2r'} \,,
\end{align}
which vanishes at $r=2\lambdabar m$. 

\subsection{Resolution of the singularity}

Let us study more in detail the boundary $r=2\lambdabar m$. We can see that curvature scalars converge to finite values there. For instance, the Ricci scalar,
\begin{align}\label{eq.riccidust}
    \mathcal{R}= \frac{2m'}{r^2r'}+\frac{6\lambdabar m^2}{r^4}\left(3+\lambdabar-\frac{16\lambdabar m}{3r}\right) +\frac{2\lambdabar m'}{r^2r'}\big(5+3\kappa\big)-\frac{12\lambdabar m m'}{r^3r'}\left(1+3\lambdabar-\frac{4\lambdabar m}{r}\right),
\end{align}
with
\begin{align}\label{eq.riccidustr0}
    \mathcal{R}\big|_{r=2\lambdabar m}=\frac{9-5\lambdabar}{8m^2\lambdabar^3}.
\end{align}
To obtain this result, we use that $r=2\lambdabar m$ corresponds to the roots of $\cos(\lambda\kang)$, and, from \eqref{eq.defm}, it follows that $r'=2\lambdabar m'$ at $r=2\lambdabar m$. Besides, Eq.~\eqref{eq.kdot2} can be analytically solved for $\kappa|_{r=2\lambdabar m}=(1/\lambdabar-1)\sinh^{-2}(y)$, with $y=\epsilon(\tau-\tau_0)\sqrt{1-\lambdabar}/(2 m\lambdabar^{3/2})$, which vanishes as $|\tau|\to\infty$, that is, as $r\to2\lambdabar m$. Finally, the convergence as $m\to0$ must follow the same regularity conditions as in GR (the origin of spherical coordinates must be regular at least for a finite interval of time in the past), in which case the curvature is regular for any $r$. Let us recall that the possible shell-crossing singularities $r'=0$ will be studied elsewhere.

Now, let the hypersurface-orthogonal geodesic vector field $\partial_\tau=\delta^\mu_\tau\partial_\mu$, and the Raychaudhuri scalar,
\begin{align}\label{eq.strongec}
    \mathcal{R}_{\mu\nu}\delta^\mu_\tau\delta^\nu_\tau=\frac{m'}{r^2r'}\left((1-\lambdabar)\left(1-\frac{10\lambdabar m}{r}\right)-3\lambdabar\kappa\right)-\frac{6\lambdabar m\kappa}{r^3}\left(1-\frac{2\lambdabar m}{r}\right)-(1-\lambdabar)\frac{\lambdabar m^2}{r^4}\left(13-\frac{24\lambdabar m}{r}\right).
\end{align}
The theory does not satisfy the timelike convergence condition (strong energy condition) of non-negative \eqref{eq.strongec}, except in the GR limit $\lambdabar=0$, and assuming $m'/r'>0$. In particular,
\begin{align}
     \mathcal{R}_{\mu\nu}\delta^\mu_\tau\delta^\nu_\tau\big|_{r=2\lambdabar m}\leq\frac{9(\lambdabar-1)}{16\lambdabar^3m^2}<0,
\end{align}
where we used $\kappa|_{r=2\lambdabar m}=0$. As a result, the focusing theorem does not apply. Finally, let us point out that the expansion scalar tends asymptotically to a constant value at $r=2\lambdabar m$,
\begin{align}
    \nabla_\mu\delta^\mu_\tau\big|_{r=2\lambdabar m}=\frac{3\epsilon}{4\lambdabar m}\sqrt{\frac{1}{\lambdabar}-1}&&\mathrm{and}&& \frac{d^n}{d\tau^n}\nabla_\mu\delta^\mu_\tau\big|_{r=2\lambdabar m}= 0,
\end{align}
for any $n\in\mathbb{Z}^+$. Its sign is just $\epsilon$, that is, the expansion scalar of $\partial_\tau$ tends eventually to a positive value in the collapse. 

To uncover the causal nature of the boundary, we need the fiducial metric $\overline{g}=\Omega^2g$. The conformal factor $\Omega=\sqrt{1-2\lambdabar m/r}$ vanishes at the desired surface, and the norm of its gradient is negative,
\begin{align}
(\overline{\nabla}_\mu\Omega)(\overline{\nabla}^\mu\Omega)|_{r=2\lambdabar m} \leq\frac{\lambdabar-1}{16\lambdabar^3m^2}<0.
\end{align}
Therefore, the nontraversable boundary $r=2\lambdabar m$ is spacelike. 

Finally, we want to stress that GR is recovered when $\lambdabar\to0$, which reduces the above to the LTB geometry, and the proper time to reach the singularity $r=0$, where the energy density and the curvature diverge, becomes finite.

\subsection{The apparent horizon}

There are yet two additional surfaces of special character, which are attainable in finite time. One of them only exists for negative $\kappa$. This is $r=2m(\lambdabar-1)/\kappa$, which describes a maximum of $r$ and a minimum of $\kappa$, because $\dot{r}=0=\dot{\kappa}$, $\ddot{r}<0$, and $\ddot{\kappa}>0$ there. The second one is $r=2m$, which is lightlike. To see it, we solve the Hamiltonian equations \eqref{eq.dotn} in a different gauge [recall that we are using \eqref{eq.defomega}]. On the way, we will show that both line elements are related by a coordinate transformation, explicitly showing the covariance of the model. On this occasion, we choose the area-radius function to be the radial coordinate, $r'=1$. We label $t$ as $T$ and, by a small abuse of notation, $x$ as $r$. Then, we can read $N^x=\epsilon\sqrt{\kappa+(1-\lambdabar)2m/r}(1-2\lambdabar m/r)$ 
from \eqref{eq.rdotn}. The metric in this new chart $\{T,r\}$ is
\begin{align}\label{eq.metricGP2}
    ds^2=\left(1+\kappa-\frac{2\lambdabar m}{r}\right)^{-1}\Bigg[&\,   -\left(1-\frac{2m}{r}\right){{dT}^{2}}+{2\epsilon}{\sqrt{\kappa+(1-\lambdabar)\frac{2m}{r}}}\left(1-\frac{2\lambdabar m}{r}\right)^{-1}{{dT} {dr}}\nonumber\\
    &\,  +\left(1-\frac{2\lambdabar m}{r}\right)^{-2}{{dr}^{2}}\Bigg]
   +r^{2}{d{\sigma}}^{2}, 
\end{align}
with $m=m(T,r)$ and $\kappa=\kappa(T,r)$ satisfying,
\begin{subequations}\label{eq.dot3}
\begin{align}
    \label{eq.mdot3}   \dot{m}&=\epsilon\sqrt{\kappa+(1-\lambdabar)\frac{2m}{r}} \left(1-\frac{2\lambdabar m}{r}\right)m',\\
    \label{eq.kdot3}\dot{\kappa}&=\epsilon\sqrt{\kappa+(1-\lambdabar)\frac{2m}{r}} \Bigg[\!\left(1-\frac{2\lambdabar m}{r}\right)\kappa'- \frac{2 \lambdabar m}{r^2}\kappa \Bigg].
    \end{align}
\end{subequations}
One can check that both line elements \eqref{eq.metricdiag2} and \eqref{eq.metricGP2} are related through the coordinate transformation $dT=d\tau$ and $dr=\dot{r}d\tau+r'dx$, with $\dot{r}=-\epsilon\sqrt{\kappa+(1-\lambdabar)2m/r}(1-2\lambdabar m/r)$. Here, we can see that $r=2m$ is a null hypersurface. 
Besides, the mean curvature vector, 
\begin{align}
    H^\mu\partial_\mu:=\frac{2}{r}(\nabla^\mu r) \partial_\mu = \frac{2}{r}\left(1-\frac{2\lambdabar m}{r}\right)\left(\epsilon\sqrt{\kappa+(1-\lambdabar)\frac{2m}{r}}\partial_T+\left(1-\frac{2\lambdabar m}{r}\right)\left(1-\frac{2 m}{r}\right)\partial_r\right),
\end{align}
with norm
\begin{align}\label{eq.meancurvaturedust}
  H_\mu H^\mu &=\frac{4}{r^2}\left(1-\frac{2m}{r}\right)\left(1-\frac{2\lambdabar m}{r}\right)^{2},
\end{align}
is also null there, and it is thus a marginally trapped surface. Then, $r=2m$ is a trapping horizon, where all the layers of matter [parametrized by a constant $x$ in the chart \eqref{eq.metricdiag2}] arrive in finite proper time.

\section{The exterior of the star}\label{sec.vac}

We have shown that the effective dust solution describes a nonending collapse. Here we turn our attention to the geometry that it may leave behind (the feasibility of the matching between both solutions is proven in Sec.~\ref{sec.matching}). 

As specified above (see also Appendix~\ref{app.extra}), the vacuum solution is very similar to the one described in Refs.~\cite{Alonso-Bardaji:2021yls,Alonso-Bardaji:2022ear}, where the Schwarzschild singularity is replaced with a totally geodesic spacelike surface of finite curvature (foliated by spheres of minimum area) in the trapped region of the spacetime. In fact, the only difference with the present model is the conformal factor $\Omega^{-2}$ in the Lorentzian sector, but this is enough to make this novel surface, which is traversable and leads to a time-reversed region in \cite{Alonso-Bardaji:2021yls,Alonso-Bardaji:2022ear}, unreachable in finite proper time. 

To study the global structure of the spacetime it is convenient to work with a nondiagonal chart. We set \eqref{eq.defomega}, and following \cite{Alonso-Bardaji:2022ear}, we will look for the most general time-independent solution. We impose $\dot{\erad}=0$ and $\dot{\ephi}=0$. Besides, we assume that $\erad'$ does not vanish identically. We solve the diffeomorphism constraint \eqref{eq.polhamdiff} for $\krad=\ephi\kang'/\erad'$, and the constant of motion \eqref{eq.defm} for $\cos(\lambda\kang)$. Let us assume that $\sin(\lambda\kang)$ does not vanish identically [the solution $\sin(\lambda\kang)=0$ yields the diagonal chart \eqref{eq.metricdiagvac}]. Then, the lapse and the shift read
\begin{align}
  \label{eq.lapsevac}      N&=\frac{c}{2}\frac{\erad'}{\ephi}\left(1-\frac{2\lambdabar m}{\sqrt{\erad}}\right)^{-1/2},\\
 \label{eq.shiftvac}       N^x&=\epsilon c\frac{\sqrt{\erad}}{\ephi}\sqrt{\frac{2m}{\sqrt{\erad}}-1+\left(\frac{\erad'}{2\ephi}\right)^{\!2}}\sqrt{1-\frac{2\lambdabar m}{\sqrt{\erad}}},
\end{align}
where $\epsilon:=-\mathrm{sgn}\big(\lambda\sin(2\lambda\kang)\big)$. The constant $c$ stands for a trivial rescaling of time, and we shall set it to one. With this, the four equations of motion are satisfied, and both constraints vanish. Hence, we find a family of metrics parametrized by $r:=\sqrt{\erad}$ and $s:=\ephi$ as functions of the radial coordinate $x$. Their specific form is pure gauge, as they can be changed through coordinate transformations. More precisely, the metric is
\begin{align}
    ds^2=\left(1-\frac{2\lambdabar m}{r(x)}\right)^{-1}\bigg[&\,-\left(1-\frac{2m}{r(x)}\right)dt^2    +\frac{2\epsilon s(x)}{\sqrt{r(x)}{\sqrt{r(x)-2\lambdabar m}}}\sqrt{\frac{2m}{r(x)}-1+\frac{r(x)^2}{s(x)^2}\big(r'(x)\big)^2}dtdx\nonumber\\
    &+\left(1-\frac{2\lambdabar m}{r(x)}\right)^{-1}\frac{s(x)^2}{r(x)^2}dx^2\bigg]+r(x)^2d\sigma^2,
\end{align}
which, as commented at the beginning of the section, is just the result (39) in Ref.~\cite{Alonso-Bardaji:2022ear} multiplied by the additional conformal factor $\Omega^{-2}=r/(r-2\lambdabar m)$ only in the Lorentzian $(t,x)$ part. This shows that there is no time-independent line element that covers $r=2\lambdabar m$.

Let us give a specific solution. We choose $r=x$ and $s^2=r^3/(r-2\lambdabar m)$. If we relabel $t$ as the function $T$ on the manifold, we find the chart $\{T,r\}$, with 
\begin{align}\label{eq.metricGPvac}
    ds^2&=
    -\left(1-\frac{2m}{r}\right)\left(1-\frac{2\lambdabar m}{r}\right)^{-1}{{dT}^{2}}+{2\epsilon}\sqrt{1-\lambdabar}\sqrt{\frac{2m}{r}}\left(1-\frac{2\lambdabar m}{r}\right)^{-2}{{dT} {dr}}+\left(1-\frac{2\lambdabar m}{r}\right)^{-3}{dr}^{2}
   +r^{2}{d{\sigma}}^{2}, 
\end{align}
which is valid for $r>2\lambdabar m$. Note that in this gauge 
\begin{align}
     N=1&&\mathrm{and}&&N^x=\epsilon\sqrt{(1-\lambdabar)\frac{2m}{r}}\left(1-\frac{2\lambdabar m}{r}\right)   , 
\end{align}
as we can see by direct substitution in \eqref{eq.lapsevac} and \eqref{eq.shiftvac}. Since the surfaces $r=2\lambdabar m$ are not covered by the chart, $\epsilon=\pm1$ is constant over the whole domain. Let us point out that this line element equals \eqref{eq.metricGP2} with zero $\kappa$ and constant~$m$.

One can easily check that the change of coordinates defined by $T=t+a(r)$, with
\begin{align}
    a(r)&=\frac{2\epsilon}{\sqrt{{1-\lambdabar}}}\Bigg((1-\lambdabar)\sqrt{2mr}+{m}\log\left(\frac{\sqrt{r}-\sqrt{2m}}{\sqrt{r}+\sqrt{2m}}\right)    -m\lambdabar^{3/2}\log\left(\frac{\sqrt{r}-\sqrt{2\lambdabar m}}{\sqrt{r}+\sqrt{2\lambdabar m}}\right)\Bigg),
\end{align}
renders \eqref{eq.metricGPvac} into \eqref{eq.metricdiagvac} [with $\Omega$ as given in \eqref{eq.defomega}]. This change of coordinates is defined only for $r>2m$, showing the isometry between this subdomain and that of \eqref{eq.metricdiagvac}.

Note that in all the above derivation, the limit $\lambdabar\to0$ yields the Schwarzschild spacetime [the usual Schwarzschild coordinates in \eqref{eq.metricdiagvac} and the Gullstrand-Painlev\'e chart in \eqref{eq.metricGPvac}]. We finish with a remark concerning the case $m\to0$ of the above geometry. This limit is the same as in Schwarzschild, that is, Minkowski. Since this is true for any value of $\lambdabar$, flat spacetime is also a solution of the effective model.

\subsection{Radial geodesics}

To completely characterize the geometry, we need to study the (causal) geodesics of \eqref{eq.metricGPvac}. Let us focus on radial geodesics parametrized as $\{T(s),r(s)\}$ with affine parameter $s$. Timelike and lightlike geodesics are determined by
\begin{align}
    \gamma&=-\left(1-\frac{2m}{r}\right)\left(1-\frac{2\lambdabar m}{r}\right)^{-1}{\left(\frac{dT}{ds}\right)^{2}}    +{2\epsilon}\sqrt{1-\lambdabar}\sqrt{\frac{2m}{r}}\left(1-\frac{2\lambdabar m}{r}\right)^{-2}{\frac{dT}{ds} \frac{dr}{ds}} +\left(1-\frac{2\lambdabar m}{r}\right)^{-3}\left(\frac{dr}{ds}\right)^{2},
\end{align}
with $\gamma=-1$ and $\gamma=0$, respectively. Besides, the conserved quantity associated with the Killing vector field $\partial_T$ is 
\begin{align}
    \mathcal{E}&:= -\left(1-\frac{2m}{r}\right)\left(1-\frac{2\lambdabar m}{r}\right)^{-1}\frac{dT}{ds}     + \epsilon\sqrt{1-\lambdabar}\sqrt{\frac{2m}{r}}\left(1-\frac{2\lambdabar m}{r}\right)^{-2}\frac{dr}{ds},
\end{align}
and we can combine both equations to get
\begin{align}\label{eq.geodesic+}
    \frac{dr}{ds}=\pm\left(1-\frac{2\lambdabar m}{r}\right)\sqrt{\left(1-\frac{2\lambdabar m}{r}\right)\mathcal{E}^2+\gamma\left(1-\frac{2 m}{r}\right)},
\end{align}
from where we can readily see that $r=2\lambdabar m$ will be achieved only after an infinite amount of affine parameter. 

Let us begin with null geodesics ($\gamma=0$). On the one hand, if $\mathcal{E}=0$, the geodesics lie on $\hor$ because $dT/ds$ cannot be vanishing when $dr/ds$ is identically zero. On the other hand, for $\mathcal{E}\neq0$, we can always choose $s$ so that $|\mathcal{E}|=1$, and we find that the affine distance diverges as $r\to\infty$, and also as $r\to2\lambdabar m$:
\begin{align}
   \pm( s&\, -s_0 )= \int \left(1-\frac{2\lambdabar m}{r}\right)^{-3/2}dr   = \frac{\sqrt{r}(r-6\lambdabar m)}{\sqrt{r-2\lambdabar m}} +{3\lambdabar m} \log\left(\frac{\sqrt{r}+\sqrt{r-2\lambdabar m}}{\sqrt{r}-\sqrt{r-2\lambdabar m}}\right).
\end{align}
For timelike geodesics ($\gamma=-1$), we have to distinguish the cases $\mathcal{E}^2=1$ and $\mathcal{E}^2>1$. We get
\begin{subequations}
    \begin{align}
  \label{eq.affinedistance} \pm( s&\, -s_0 )= \frac{2}{3}\sqrt{\frac{r}{2m}}\frac{r+6\lambdabar m}{\sqrt{1-\lambdabar}}  -\frac{2m \lambdabar^{3/2}}{\sqrt{1-\lambdabar}}\log\left(\frac{\sqrt{r}+\sqrt{2\lambdabar m}}{\sqrt{r}-\sqrt{2\lambdabar m}}\right),\\
    \pm( s&\, -s_0 )= \frac{\sqrt{r}\sqrt{f}}{\mathcal{E}^2-1}    -\frac{2m\lambdabar^{3/2}}{\sqrt{1-\lambda}}\log\left|\frac{\sqrt{r}\sqrt{1-\lambdabar}+\sqrt{\lambdabar}\sqrt{f}}{\sqrt{r}\sqrt{1-\lambdabar}-\sqrt{\lambdabar}\sqrt{f}}\right|    -\frac{1+(2-3\mathcal{E}^2)\lambdabar}{(\mathcal{E}^2-1)^{3/2}}\log\left|\frac{\sqrt{f}+\sqrt{r}\sqrt{\mathcal{E}^2-1}}{\sqrt{f}-\sqrt{r}\sqrt{\mathcal{E}^2-1}}\right|,
\end{align}
\end{subequations}
respectively, with $f:=\left(r-{2\lambdabar m}\right)\mathcal{E}^2-\left(r-{2 m}\right)\geq0$. In both cases, the right-hand side goes to $+\infty$ as $r\to\infty$. The novel second term (which vanishes for $\lambdabar=0$) also diverges when $r\to2\lambdabar m$ (in this case, the right-hand sides of the equalities tend to $-\infty$). 

As a result, both lightlike and timelike radial geodesics can be extended to infinite values of the affine parameter in both directions (increasing and decreasing area of the spheres). Thus, $r=2\lambdabar m$ emerges as an asymptotic boundary of spacetime located at a finite value of the area-radius function.

\subsection{Global structure}\label{sec.globalvac}

Let us define the relevant (nonoverlapping) subsets of the domain described by the chart \eqref{eq.metricGPvac}: $E:=\{r>2m\}$, $\hor:=\{r=2m\}$, and $I:=\{2\lambdabar m<r<2m\}$. Let us also split these sets under the sign of $\epsilon$, and denote them as $D^\epsilon$, for $D=E,\hor,I$. The surfaces of constant $r$ are timelike, lightlike, and spacelike, respectively, in each of the subdomains. In addition, the spheres $r>2m$ are nontrapped, the surfaces $r=2m$ are marginally trapped, and the spheres $2\lambdabar m<r<2m$ are trapped to the future (past) when $\epsilon=+1$ ($\epsilon=-1$). To check this, we need the mean curvature vector,
\begin{align}
    H^\mu\partial_\mu:=\frac{2}{r}(\nabla^\mu r)\partial_\mu=\frac{2}{r}\left(1-\frac{2\lambdabar m}{r}\right)\left(\epsilon\sqrt{(1-\lambdabar)\frac{2m}{r}}\partial_T+\left(1-\frac{2\lambdabar m}{r}\right)\left(1-\frac{2m}{r}\right)\partial_r\right),
\end{align}
and a future-pointing unit normal, i.e., $n_\mu dx^\mu=- dT$ in the chart \eqref{eq.metricGPvac}. Then, 
\begin{subequations}\label{eq.meancurvature}
\begin{align}
  \label{eq.normgradr}  H_\mu H^\mu &=\frac{4}{r^2}\left(1-\frac{2m}{r}\right)\left(1-\frac{2\lambdabar m}{r}\right)^{2},\\
    n_\mu H^\mu &=-\epsilon\sqrt{(1-\lambdabar)\frac{8m}{r^3}}\left(1-\frac{2\lambdabar m}{r}\right),
\end{align}
\end{subequations}
showing that the mean curvature vector is spacelike at $E^\pm$, lightlike at $\hor^\pm$, and future-pointing (past-pointing) timelike at $I^+$ ($I^-$). Finally, we also need to define the sets $\mathfrak{T}^\pm:=\{r=2\lambdabar m\}\cap\{\epsilon=\pm1\}$, which are future (past) boundaries of $I^+$ ($I^-$).

The global picture of the spacetime solution is clear, at least up to the nature of these boundaries. Let $\overline{g}=\Omega^2g$ be a fiducial metric. The conformal factor $\Omega=\sqrt{1-2\lambdabar m/r}$, with $d\Omega=\lambdabar m/(r^2\Omega)dr$, satisfies $\Omega|_{r=2\lambdabar m}=0$ and
\begin{align}
    (\overline{\nabla}_\mu\Omega)(\overline{\nabla}^\mu\Omega)\big|_{r=2\lambdabar m} =\frac{\lambdabar-1}{16\lambdabar^3m^2}<0,
\end{align}
so $\mathfrak{T}^\pm$ are spacelike. In fact, since $\overline{g}$ is the geometry studied in Refs.~\cite{Alonso-Bardaji:2022ear}, and the additional conformal factor cannot change the causal structure, we can readily import the construction of the Penrose diagram from that reference. Since the only difference with those references lies on the fact that $r=2\lambdabar m$ is no longer traversable, not only $E=E^+\cup E^-$ but also $I=I^+\cup I^-$ are disconnected domains. 

\begin{figure}
    \centering
    \includegraphics[width=.8\textwidth]{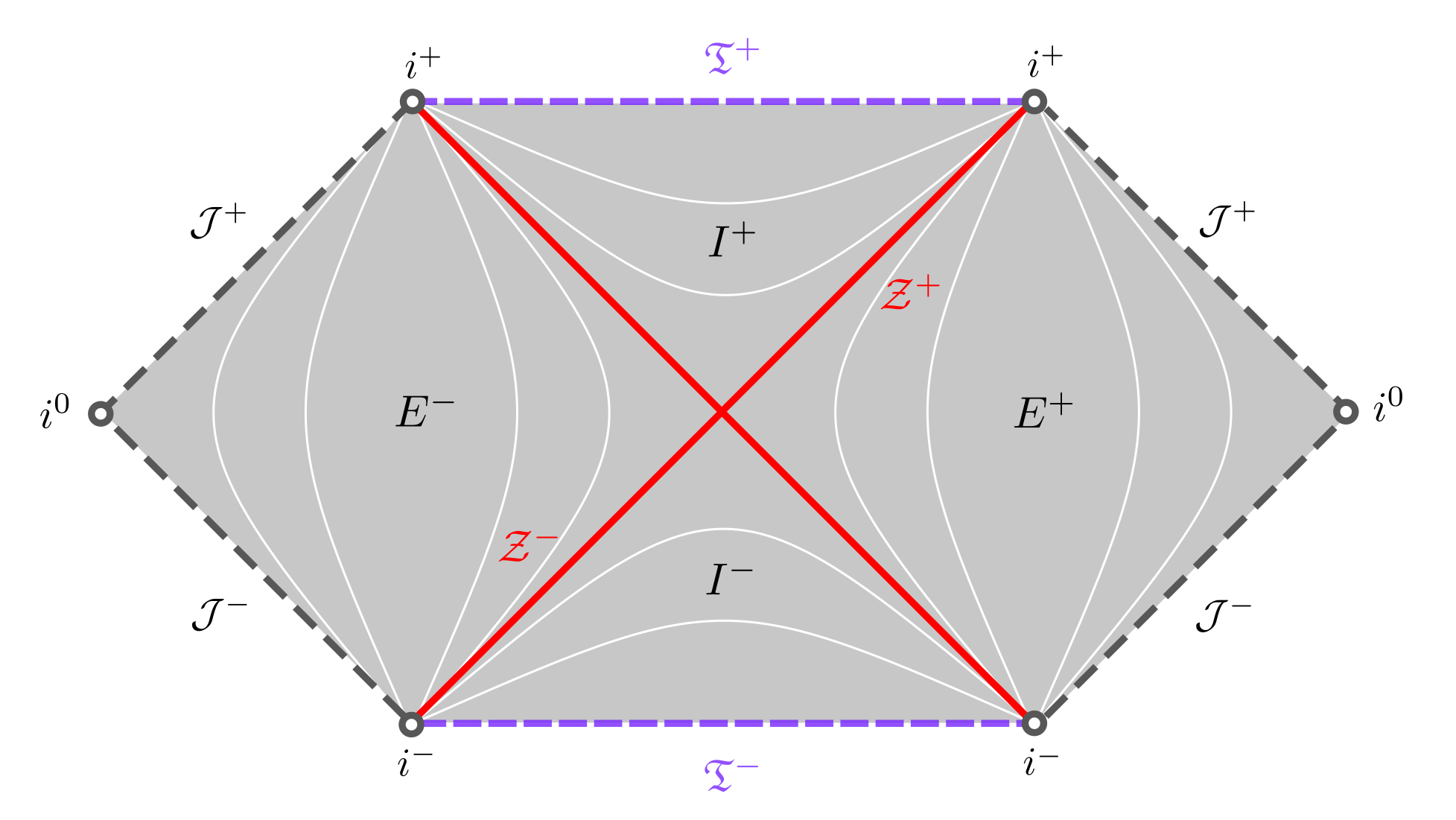}
    \caption{Penrose diagram of the maximally extended solution. The chart \eqref{eq.metricGPvac} covers either of the diagonals. White lines are those of constant $r$, and we draw in red the black-hole horizon. Dashed lines represent the $\mathbb{R}\times S^2$ boundaries of the spacetime. In gray the usual null asymptotic infinities, and in purple the novel boundaries (of finite curvature) replacing the singularity. Gray rings are timelike and spacelike infinities.}
    \label{fig.vacuum}
\end{figure}

For completeness, we present the compactified chart,
\begin{align}
    ds^2=\frac{\widetilde{\Gamma}(r(u,v))}{\cos^2u\cos^2v} du dv +r(u,v)^2d\sigma^2,
\end{align}
with the everywhere negative
\begin{align}
    \widetilde{\Gamma}(r):=&\,-\frac{32m^3}{(1-\lambdabar)(r-2\lambdabar m)}\left(\sqrt{1-\lambdabar}+\sqrt{1-\frac{2\lambdabar m}{r}}\right)^2\exp\left[-\sqrt{1-\lambdabar}\sqrt{1-\frac{2\lambdabar m}{r}}\frac{r}{2m}\right]\nonumber\\
    &\,\times \left(1+\sqrt{1-\frac{2\lambdabar m}{r}}\right)^{-\sqrt{1-\lambdabar}(2+\lambdabar)}\left(\frac{2\lambdabar m}{r}\right)^{\sqrt{1-\lambdabar}(1+\lambdabar/2)-1},
\end{align}
and $r=r(u,v)$ implicitly defined through
\begin{align}
    \tan u\tan v =\left(1-\frac{2m}{r}\right)\frac{16m^2}{(1-\lambdabar)\widetilde{\Gamma}(r)},
\end{align}
which is valid on the open region $\{|u|<\pi/2,|v|<\pi/2,|u+v|<\pi/2\}$. Note that $u=0$ and $v=0$ correspond clearly to $r=2m$. 
In Fig.~\ref{fig.vacuum} we draw the maximal extension of this new spacetime. The chart \eqref{eq.metricGPvac} covers any of the diagonals, $E^+\cup\hor^+\cup I^+$ or $I^-\cup\hor^-\cup E^-$ (and, by isometry, $E^-\cup\hor^+\cup I^+$ or $I^-\cup\hor^-\cup E^+$), for $\epsilon=+1$ and $\epsilon=-1$, respectively. 
In contrast to previous studies, there is no bounce to a time-reversed (white-hole) region of spacetime, and any observer falling inside the black hole is doomed to travel {forever} towards the hypersurface $\mathfrak{T}^+$, foliated by spheres of infimum area $\Delta:=16\pi \lambdabar^2 m^2$. In the GR limit $\lambdabar\to0$, we find $\Gamma(r)\to -32 m^3/r \exp[-r/2m]$, which is just the Kruskal-Szekeres coefficient of the maximally extended Schwarzschild spacetime.

\begin{figure}
    \centering
    \includegraphics[width=0.75\textwidth]{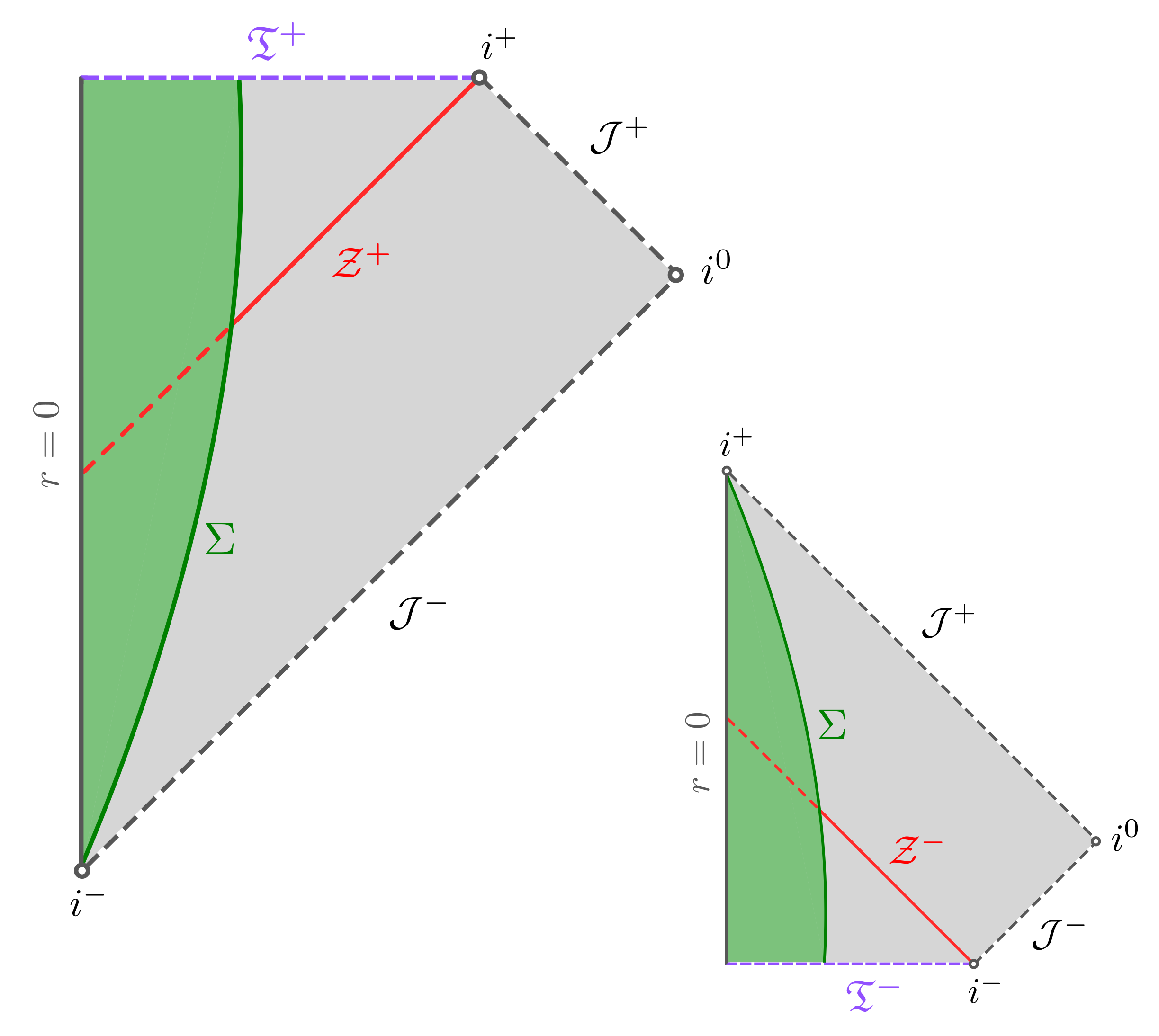}
    \caption{Top left: Penrose diagram of the full spacetime describing the effective collapse of a spherical dust star. In green the ``internal'' matter solution, with the dark green line being the surface of the star (the matching hypersurface $\Sigma$). In gray, the ``external'' vacuum solution, where the red line $\mathcal{Z}^+$ represents the event horizon. The dashed purple line $\mathfrak{T}^+$ is the novel hypersurface that replaces the singularity. All curvature scalars are finite there, and infalling particles travel towards it without ever reaching it. The spheres within this boundary attain the infimum area $\Delta:=16\pi\lambdabar^2m^2$. In dashed gray, the asymptotic null infinities. The gray rings represent timelike and spacelike infinities. Bottom right: The time reversed solution.}
     \label{fig.complete}
\end{figure}

The case under study in this work is the nonsingular collapse of a spherical star. To describe the complete spacetime we thus need to identify the matter solution as the internal part of the matching problem performed in Sec.~\ref{sec.matching}. Then, the vacuum will be the external region. The full Penrose diagram for the collapsing matter (green) surrounded by the vacuum solution (gray) is represented in Fig.~\ref{fig.complete}. Note that since the mass parameter is equal from both sides at $\Sigma$, there are no discontinuities in the apparent horizon. The time-reversed solution is just a mirror symmetry around the axis $\mathfrak{T}$, inverting the $+$ and $-$ signs in the diagram. Let us remark that the complementary matching with the discarded pieces of the matter and vacuum solutions is also possible under the same conditions. In the limit $\lambdabar\to0$, the boundaries $\mathfrak{T}$ emerge as the GR singularities.

\subsection{Curvature}\label{sec.curvature}

Finally, let us show that, in contrast to $r=0$ in GR, the curvature is well behaved in a neighborhood of $\mathfrak{T}^\pm$. In fact, the relevant curvature scalars, the Ricci scalar and the only nonvanishing component of the Weyl tensor, 
\begin{align}
  \label{eq.ricci}  \mathcal{R}&=\frac{6\lambdabar m^2}{r^4}\left(3+\lambdabar-\frac{16\lambdabar m}{3r}\right),\\
    \Psi_2&=-\frac{m}{r^3}\left(1+\frac{\lambdabar}{2}\right)+\frac{\lambdabar m^2}{2r^4}\left(7+\lambdabar\right)-\frac{10\lambdabar^2m^3}{3r^5},
\end{align}
respectively, are everywhere finite. Note that in the Minkowski limit $m=0$ both scalars are exactly zero. For $m\neq0$, 
\begin{align}\label{eq.bounds}
    \mathcal{R}\big|_{r=2\lambdabar m}&=\frac{1+3\lambdabar}{8m^2\lambdabar^3},\\
    \Psi_2\big|_{r=2\lambdabar m}&=-\frac{1+3\lambdabar}{96m^2\lambdabar^3}.
\end{align}
Note that the former is always smaller than its counterpart in the matter solution \eqref{eq.riccidustr0} because $\lambdabar<1$.  

The two-dimensional Lorentzian manifold has a constant curvature ${}^{(2)}\mathcal{R}=(1-\lambdabar)/(8m^2\lambdabar^3)$ when approaching $\mathfrak{T}^\pm$, and it is thus asymptotically de Sitter, with metric $ds_{(2)}^2|_{\mathfrak{T}}=-dT^2+\exp[2{T}/\alpha]dX^2$ with $\alpha:={4m\lambdabar^{3/2}}/\sqrt{{1}-{\lambdabar}}$, while the metric on the two-sphere is just $4\lambdabar^2m^2d\sigma^2$. 

For completeness, we also include the Kretschmann scalar,
\begin{align}
     \mathcal{K}&=\frac{48 m^2}{r^6}+\frac{24\lambdabar m^2}{r^6}\left(2+5 \lambdabar\right)-\frac{48 \lambdabar m^3}{r^7}\left(7+10\lambdabar+7 \lambdabar^2\right) \nonumber\\&\quad
     +\frac{4 \lambdabar^2 m^4}{r^8}\left(289+318 \lambdabar +65 \lambdabar^2\right) 
     -\frac{ 1024 \lambdabar^3 m^5}{r^9}(2+\lambdabar)+\frac{1408 \lambdabar^4 m^6}{r^{10}} ,
\end{align}
with
\begin{align}\label{eq.maxK}
    \mathcal{K}\big|_{r=2\lambdabar m}=\frac{1-2\lambdabar+17\lambdabar^2}{64m^4\lambdabar^6}.
\end{align}
Then, the curvature is well behaved in a neighborhood of $\mathfrak{T}^\pm$, and it converges to finite values.

{As we commented in the introduction, this geometry resembles that found in Refs.~\cite{Han:2022rsx,Han:2020uhb}. While both resolve singularities by means of a finite-curvature spacelike boundary, there are relevant differences in the construction of the models. The curvature in this novel surface also differs: It goes as $\Delta^{-2}$ in the mentioned references while \eqref{eq.maxK} depends on the black-hole size, i.e., the terms $ m^2/\Delta^3$, $m/\Delta^{5/2}$, and $1/\Delta^2$ are present in this limit. More fundamentally, the studies can be related to mimetic-dilaton-gravity models (so they descend from a covariant Lagrangian), they implement an improved-dynamics scheme (to avoid quantum effects at low curvatures) in both curvature components and work within a reduced-phase-space framework, where the dust field plays the role of a reference clock. The additional complexity of the Hamiltonian makes it tracktable only numerically, and it is not possible to go outside the dust foliation. In contrast, we ensure explicit covariance in the Hamiltonian, avoiding ambiguities in defining the spacetime metric. Although it is probably less motivated by LQG, we implement a constant-polymerization scheme only in the angular component of the extrinsic curvature. This allows us to build exact analytical solutions for both the vacuum and matter solutions. Besides, we do not find deviations from GR at large scales: relevant quantum effects are confined to the trapped region because $\lambdabar<1$. Nevertheless, they might become dominant in some regimes, as we study in the following via the black-hole evaporation process. Let us remark that while the methods and assumptions differ significantly, both approaches provide a compatible effective description of LQG, contributing in our understanding of quantum-black-hole physics.}

\section{Black-Hole Thermodynamics and Remnants}\label{sec.evaporation}

Up to this point, the main focus has been on $\lambdabar$ being constant over the whole phase space. The main advantage of this viewpoint is that the limit $m\to0$ of the above family of metrics is Minkowski, independently of the value of the polymerization parameter. Therefore, flat spacetime is a solution of the effective theory. This consideration corresponds to the $\mu_0$-scheme in loop quantum cosmology. However, it shows some unexpected properties. For instance, the infimum area $\Delta:=16\pi \lambdabar^2 m^2$ increases with the mass, and the upper bounds of the curvature diminish accordingly (see Sec.~\ref{sec.curvature}). To face the problems of a constant polymerization, there have been many attempts to implement a scale-dependent polymerization, the so-called $\bar{\mu}$-scheme, in loop quantum cosmology. While it is unclear whether this actually improves the spherical models \cite{Ashtekar_2018}, in vacuum there is yet a third option: Consider a scale-independent polymerization parameter, but in this case being a functional of the (constant) mass \cite{Corichi:2015xia,Olmedo:2017lvt}. This promotes the infimum area of the spheres, $\Delta$, to the fundamental constant of the theory, and $\lambdabar=\sqrt{{\Delta}/{(16\pi)}}\,m^{-1}$ is constant over any dynamical trajectory (without altering the equations of motion) but varies for each different solution (just as the parameter $m$ in Schwarzschild). In this way, the curvature bounds increase with the mass: The ``heavier'', the more curved. For instance, the Ricci scalar satisfies $\mathcal{R}|_{\mathfrak{T}}=6\pi/\Delta+8m(\pi/\Delta)^{3/2}$. 

To simplify the presentation, let us define $m_0:=\sqrt{{\Delta}/{(16\pi)}}$. The limiting case $m\to m_0$ corresponds to $\lambdabar\to1$, i.e., a divergent polymerization parameter $\lambda\to\infty$. This shows that astrophysical black holes (of big mass) have negligible corrections parametrized by $\lambdabar=m_0/m\ll1$, while still being free of singularities. Then, quantum-gravity effects become macroscopic for masses close to the infimum, which are presumably Planckian and thus lie in the limit of validity of the model. In the following, we prove that $m=m_0$ is achieved asymptotically through (semi)classical evaporation, predicting in that way the formation of stable black-hole remnants. 

But first, let us remark that the quasilocal Hawking [see \eqref{eq.hawkingmass}] and Komar masses do not coincide as in GR \cite{BEIG1978153,Alonso-Bardaji:2022ear}, and we consider the former to describe the energy of the system,
\begin{align}\label{eq.energy}
    E:=M_H=m+{2m_0}\bigg(1-\frac{2m}{r}\bigg)\bigg(1-\frac{m_0}{r}\bigg).
\end{align}
There are technical reasons for this choice. First, the Hawking mass measures the energy contained within the spheres of constant time and radial coordinates and, in particular, within $r=2m$. Second, it is always greater than the Komar mass outside the horizon\footnote{A straightforward calculation yields
\begin{align*}
    M_K:=\frac{1}{4\pi}\int_\mathcal{S} {\hat{\xi}^\mu \hat{s}_\nu\nabla_\mu \xi^\nu}
    d\mathcal{S} = m\sqrt{1-2m_0/r},
\end{align*}
with $\hat{\xi}$ the unit timelike Killing field and $\hat{s}$ the unit outward-pointing normal to the enclosing surface $\mathcal{S}=S^2$ (which is also orthogonal to $\hat{\xi}$). Since both the Hawking and Komar masses are monotonically increasing functions of $r$, and for $r\geq2m$, the minimum of the first coincides with the maximum of the second, so $E>M_K$. The weak Dirac observable \eqref{eq.defm} coincides with the asymptotic value of the Komar mass. In addition, the surface gravity $\kappa$ is constant and satisfies $\kappa=r^{-2}M_K|_\hor$.}. Third, and perhaps most importantly, the entropy associated with the black hole vanishes in the zero-temperature case when we take the Hawking energy (see below), but it would diverge if we considered the Komar mass, as one can check by repeating the following computations with the Komar mass instead. 

We also define the asymptotic mass 
\begin{align}\label{eq.massasymptotic}
    M:=m+2m_0,
\end{align}
which is a geometric invariant and corresponds to the value of the Hawking mass at infinity.

We readily see that the parameter $m$ is related to but cannot be identified with the mass of the black hole. As shown before, it rather defines the horizon, with area $A_\hor:=16\pi m^2\equiv 16\pi M_K^2|_{i^0}$, which is clearly smaller than $16\pi M^2\equiv16\pi E^2|_{i^0}$. Their difference is $64\pi m_0(M-m_0)$. It is straightforward to find $E|_\hor=m$ and $dE|_\hor=dm=dM$.

The maximal extension of the geometry comprises a bifurcate Killing horizon of the Killing field, completely analogous to the GR case (see Fig.~\ref{fig.vacuum}). Then, on any of the asymptotic (exterior) regions, there exists a thermal state for the quantum fields propagating on the background spacetime~\eqref{eq.metricdiagvac}. In other words, we find an induced thermal spectrum of temperature $1/(2\pi)$ times the (constant) surface gravity of the horizon. Explicitly,
\begin{align}\label{eq.temperature}
    T:=\frac{1}{2\pi}\sqrt{-\frac{1}{2}(\nabla_\mu t_\nu)(\nabla^\mu t^\nu)}\bigg|_{\hor} \!= \frac{1}{8\pi m}\sqrt{1-\frac{m_0}{m}}.
\end{align}
In clear contrast to GR, where this function decreases monotonically with the mass, and diverges as $m\to0$, the temperature \eqref{eq.temperature} has a maximum $T=1/(8\pi\sqrt{3}m_0)$ at $m=3m_0/2$, and vanishes for $m=m_0$. Note that if the polymerization parameter is the fundamental constant of the model, the temperature $T=\sqrt{1-\lambdabar}/(8\pi m)$ still diverges in the vanishing mass limit (which may be problematic even if there is no naked singularity), before giving birth to a new flat spacetime as studied in Ref.~\cite{Balart:2024rts} for the bouncing model \cite{Alonso-Bardaji:2021yls,Alonso-Bardaji:2022ear}. 

Knowing the energy and the temperature, it is easy to compute the entropy associated with the black-hole horizon through the first law of thermodynamics $dE=TdS$. With \eqref{eq.energy} and \eqref{eq.temperature}, we find
\begin{align}\label{eq.entropy}
    S&=8\pi\int \frac{m^{3/2}dm}{\sqrt{m-m_0}}=4\pi m^2\left(1+\frac{3m_0}{2m}\right)\sqrt{1-\frac{m_0}{m}}+3\pi m_0^2\log\frac{\sqrt{m}+\sqrt{m-m_0}}{\sqrt{m}-\sqrt{m-m_0}} +S_0,
\end{align}
The constant minimum entropy $S=S_0$ is attained in the vanishing temperature case $m=m_0$, as stipulated by the third law of thermodynamics. Note that the leading correction in the large-mass limit is {not} logarithmic in the horizon area, but rather
\begin{align}
    S= \frac{A_\hor}{4}+\frac{\sqrt{\Delta A_\hor}}{4}+\frac{3\Delta}{16} \log A_\hor +\widetilde{S}_0+\mathcal{O}\big(A_\hor^{-1/2}\big)
\end{align}
where we have also used $\Delta=16\pi m_0^2$ and $\widetilde{S}_0=S_0-(7-6\log4){\Delta}/{32}$. This entropy-area law is in complete accordance with previous studies in LQG where geometric excitations of the horizon are assumed indistinguishable and the degeneracy of nongeometric degrees of freedom is exponential with the area \cite{Frodden_2014,Ghosh_2014,Asin_2015,Barrau2018}. In addition, the expansion for the large area is independent of the choice of mass and conformal factor in the Lorentzian sector, because the same relation was found in Ref.~\cite{Borges_2024} for the twin model \cite{Alonso-Bardaji:2021yls,Alonso-Bardaji:2022ear}. In the GR limit, this reduces to the usual expression $S=A_{\hor}/4+{S}_0$. 

\begin{figure}
    \centering
    \includegraphics[width=0.55\textwidth]{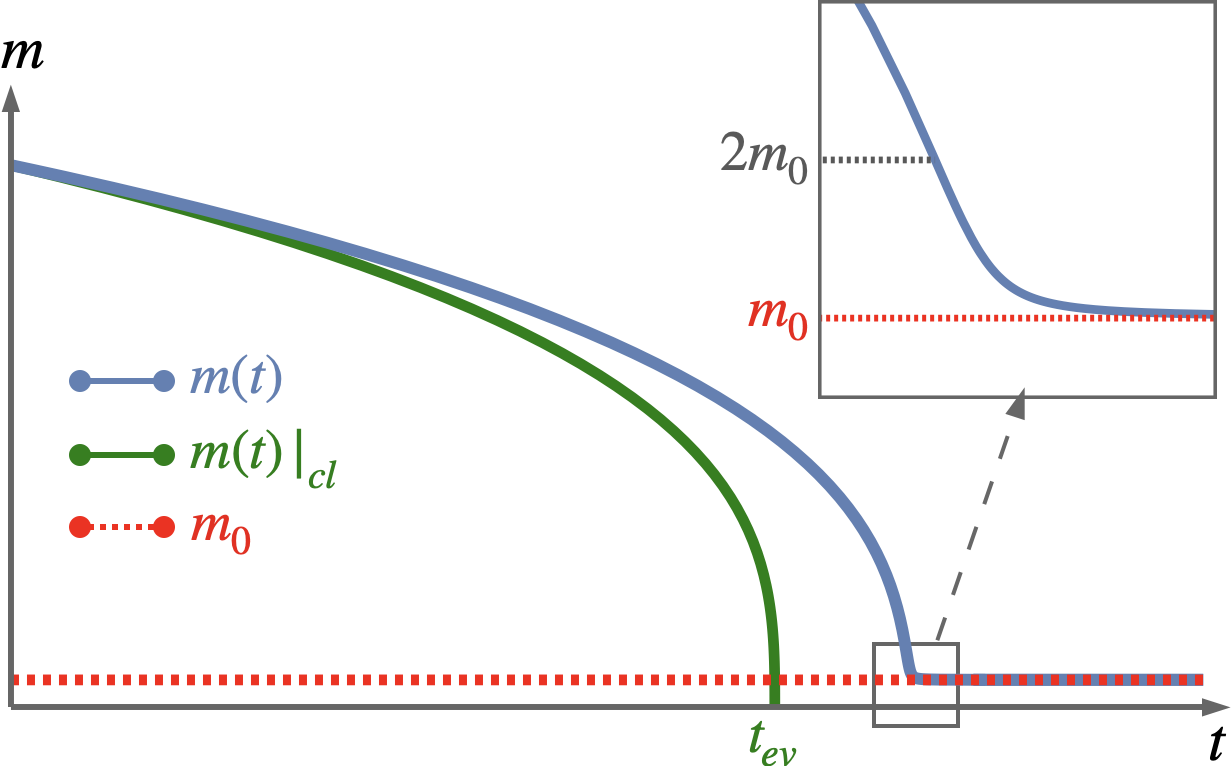}
    \caption{The function $m(t)$ in the evaporation process (blue), and its comparison with the classical trajectory (green). The behavior is qualitatively similar until very small masses. But while the classical trajectory ends at finite $t=t_{ev}$, the effective $m(t)$ tends asymptotically to the infimum $m=m_0$ (dashed red). Above, zoom of the inflection region around $m=2m_0$.}
    \label{fig.m(t)}
\end{figure}

If the black hole mimics a perfect black body, it emits thermal radiation following the Stephan-Boltzmann law, $dE/dt=-\sigma A T^4$, with $\sigma:=\pi^2/60$ in natural units and $A=A_\hor$ the area of the horizon. Let us further assume that the evaporating black hole can be approximated by the stationary geometry \eqref{eq.metricdiagvac} (with different values of the mass parameter) at every instant. Then, we can easily prove that the state of the vanishing temperature and constant minimum entropy cannot be attained in finite time. The radiated energy makes the mass of the black hole $M(t)=m(t)+2m_0$ to monotonically decrease as
\begin{align}\label{eq.evaporation}
    \frac{dm}{dt}=-\frac{1}{15360\pi m^2}\left(1-\frac{m_0}{m}\right)^2.
\end{align}
This function has an inflection point at $m=2m_0$, where the evaporation starts to decelerate, and it approaches asymptotically $m=m_0$ from above, as shown in Fig.~\ref{fig.m(t)}. We can integrate this differential equation to obtain
\begin{align}\label{eq.evaporationtime}
    &\frac{t_f-t_i}{61440\pi}=\frac{m_i^3-m_f^3}{3}-\frac{1}{4}\left(\frac{m_i^4}{m_i-m_0}-\frac{m_f^4}{m_f-m_0}\right)+\frac{m_0}{2}(m_i^2-m_f^2)+m_0^2(m_i-m_f)
    +m_0^3\log\frac{m_i-m_0}{m_f-m_0},
\end{align}
with $t_i\leq t_f$ and $m_0<m_f\leq m_i$. In contrast to GR, where the complete evaporation takes finite time, the infimum $m_f=m_0$ can be attained only after infinite time. For large masses $m_i\gg m_0$, we find that the difference with the ``classical'' time to achieve $m_f=2m_0$ is quadratic in the initial mass, i.e., ${t_f- t_f|_{cl}} \approx 10^{5}m_0m_i^2$ (and hence negligible, because $t_f|_{cl}\sim m_i^3$). The limit case $m_f=m_0$ corresponds to a black hole of zero surface gravity \eqref{eq.temperature} and constant minimum entropy \eqref{eq.entropy}, which we shall call a black-hole remnant. 

Even if the quasistatic approximation was reliable up to these scales, the continuous asymptotic limit could not be physically realistic due to the quantum nature of energy. Going back to \eqref{eq.evaporationtime}, and setting $m_f=(1+\varepsilon)m_0$, we find $t_f\sim m_0^3/\varepsilon$ and $T\sim\sqrt{\varepsilon}/m_0$ as $\varepsilon\to0$. Hence, it is reasonable that the evaporation ends up in finite time at masses slightly above the infimum. Besides, full quantum-gravity phenomena, which are clearly beyond the reach of the present work (and they cannot be described within the effective theory), may lead to spontaneous decays of the remnant; for instance, into a large number of low-energy photons~\cite{Kazemian:2022ihc}. But even if this was the case, the estimated scale for such decays is 1 order of magnitude bigger than the evaporation time (the half-life time is of the order $m_i^4$). Therefore, the remnants predicted by the effective theory are metastable at least. If their mass lies in the suitable range, they could stand for long-lived dark-matter particles.

We want to remark that the theory is model independent. That is, even if the effective singularity-free theory is motivated by LQG, the constant correction parameters can be assumed to be completely free (of course, large corrections for astrophysical scenarios can be discarded by current observations) with no consequence on the predictions thorough the work: Infimum positive area, resolution of singularities, and existence of remnants. But we now break this idea by identifying the infimum area of the spheres, $\Delta=16\pi m_0^2$, with the LQG area gap $\Delta_\gamma:=4\pi\sqrt{3}\gamma$, where $\gamma$ is the Barbero-Immirzi parameter. In this way, we can estimate the mass of such black-hole remnants within effective LQG. {More precisely, the black hole remnants formed after an infinite evaporation time, with a horizon area equal to the minimum eigenvalue of the area operator in LQG, have an estimated mass of}
\begin{align}\label{eq.massremnant}
      M_{rem}:=M\big|_{m=m_0}=\sqrt[4]{\frac{243}{16}\gamma^2}\,,
\end{align}
{within this effective theory.} 
Considering the range $\frac{\log(2)}{\pi}\leq\gamma\leq\frac{\log(3)}{\pi}$ found in Ref.~\cite{Domagala_2004}, we get $0.92<M_{rem}<1.17$, which, as one would expect, is close to (in fact includes) the Planck mass. {Therefore, the effective theory predicts the existence and the formation of some objects that lie precisely within the natural scale of quantum gravity $m_P$. Such Planckian ``particles'', which are expected to interact only gravitationally, postulate as an interesting (and possibly measurable \cite{Perez:2023tld}) candidate for dark matter \cite{MacGibbon:1987my, Giddings:1992hh, Bianchi:2018mml, Amadei_2022, Rovelli:2024sjl}. Of course, this represents a limiting case that cannot be reached through thermal emission. As stated above, it provides a lower bound for remnant masses. Moreover, stronger backreaction, full quantum-gravity effects, or the breakdown of the quasistatic approximation may become increasingly significant as this limit is approached. In other words, remnants are fully quantum-gravitational objects suggested by, but not contained within, the framework of the effective theory.} If we set the widely accepted value $\gamma=0.2375$ \cite{Meissner:2004ju}, we find 
\begin{align}\label{eq.massremnantN}
      M_{rem}\approx0.9621 m_P=20.94\mu \mathrm{g},
\end{align}
where we have reintroduced the units for clarity. {The result is surprisingly close to the one-Planck-mass value expected from quantum-gravity realizations. As mentioned above, this represents only the lower bound permitted within the effective theory, and we should remain cautious regarding the precise numerical value.} In turn, experimental bounds on the mass of these black-hole-remnant particles could help to determine the Barbero-Immirzi parameter. Relying on the above formula, and for remnant masses of exactly one Planck mass, this effective model predicts $\gamma=\sqrt{16/243}\approx0.2566$, {which still lies within the expected range $\frac{\log(2)}{\pi}\leq\gamma\leq\frac{\log(3)}{\pi}$~\cite{Domagala_2004}.}

\section{Summary}\label{sec.conclusions}

We have presented a covariant Hamiltonian \eqref{eq.hampoltot} and its associated metric \eqref{eq.metricfinal} that completely describe the nonsingular collapse of dust in effective spherical LQG. The main result of this work is that we are able to describe the whole spacetime solution of a dynamical star surrounded by vacuum. In addition, the model predicts the formation of stable black-hole remnants via thermal evaporation. {In other words, we provide a complete collapse model, from black-hole formation to remnant stabilization.} 

The effective theory motivated by LQG generalizes the LTB spacetimes \eqref{eq.metricdiag2}--\eqref{eq.dot2} by introducing a novel constant in the model: The polymerization parameter of the holonomy corrections. Besides, these minimally coupled effective dust solutions can be matched to the unique family of biparametric vacuum geometries \eqref{eq.metricGPvac} (see the conformal diagram of this vacuum spacetime in Fig.~\ref{fig.vacuum}). The matching hypersurface is comoving with the flow of matter, and we identify it with the surface of the star (see Fig.~\ref{fig.complete} for the conformal diagram of the complete spacetime solution). The matching conditions plus the absence of singularities in the matter solution fix the free function of the model. In all cases, the GR singularities are resolved, and they are replaced with a spacelike boundary of the trapped region of spacetime. This boundary is foliated by spheres of infimum positive area, and the curvature converges there to finite values. Radial timelike and null geodesics inside the black hole can be extended to infinite values of the affine parameter in both directions (increasing and decreasing area of the spheres), and the spacetime is thus free of any singularities. The absence of a geometric bounce to a time-reversed region removes all the problems associated with Cauchy horizons and white-hole instabilities. When the polymerization parameter goes to zero, spherical GR emerges as a singular limit of the theory. 

To conclude, we have studied the evaporation of these static black holes through thermal radiation. When considering the infimum of the area to be the fundamental constant of the theory (as motivated by LQG), the effective model predicts the formation of stable black-hole remnants. In complete agreement with the third law of black-hole thermodynamics, this final state of zero surface gravity and constant minimum entropy cannot be achieved in finite time. Considering the value $\gamma=0.2375$ of the Barbero-Immirzi parameter, we find that the lower bound for remnant masses is of almost one Planck mass: $20.94\mu$g. If the effective theory is reliable at these scales, it provides an interesting framework to study a promising dark-matter candidate.

\begin{acknowledgments}
    I am most grateful to David Brizuela and Ra\"ul Vera for enlightening discussions and a careful reading of this manuscript. I also thank Alejandro Perez for interesting comments. This work was supported by the Grant PID2021-123226NB-I00 (funded by MCIN/AEI/10.13039/501100011033 and by “ERDF A way of making Europe”) and by the ID\#~62312 grant from the John Templeton Foundation, as part of the project \href{https://www.templeton.org/grant/the-quantum-information-structure-of-spacetime-qiss-second-phase}{``The Quantum Information Structure of Spacetime'' (QISS)}. The opinions expressed in this work are those of the author and do not necessarily reflect the views of the John Templeton Foundation.
\end{acknowledgments}

\appendix

\section{Motivation and relation to previous works}\label{app.extra}

In this work we show that a subtle generalization of the models in Refs.~\cite{Alonso-Bardaji:2021yls,Alonso-Bardaji:2022ear,Alonso-Bardaji:2023qgu} allows to describe a nonsingular spherical collapse of a star of minimally coupled (effective) dust. In turn, the vacuum reduction of the model also describes a geodesically complete black-hole spacetime. The motivation for this generalization is as follows: The singularity resolution in \cite{Alonso-Bardaji:2021yls,Alonso-Bardaji:2022ear} is unstable under the addition of simple matter fields \cite{Bojowald:2023djr}. Nevertheless, in Ref.~\cite{Alonso-Bardaji:2023qgu}, it was shown that trajectories on phase space are completely regular and that there is an infinite family of associated metrics describing a well-behaved geometry (with no curvature divergences). The question is whether the ambiguity on the construction of the metric can be translated to the Hamiltonian. As we show in the following, the answer is positive, and a conformal factor in the two-dimensional Lorentzian submanifold can be absorbed in the Hamiltonian. This holds in vacuum, but as soon as one adds minimally coupled matter the analogy breaks, and the models are no longer equivalent (one can compare the results in this paper with those in Ref.~\cite{Alonso-Bardaji:2023qgu}). 

Let us consider \eqref{polham}, and write the total Hamiltonian,
\begin{align}
    H_T=\int \big(N\ham_g +N^x\diff_g\big)dx= \int \big(N\Omega\overline{\ham}_g +N^x\diff_g\big)dx = \int \big(\overline{N}\,\overline{\ham}_g +N^x\diff_g\big)dx,
\end{align}
where $\overline{\ham}_g:=\ham_g\big|_{\Omega=1}$ is the Hamiltonian constraint in \cite{Alonso-Bardaji:2021yls,Alonso-Bardaji:2022ear} and we have defined $\overline{N}:=\Omega N$. 

Now, let us disclose some {off-shell} properties of the scalar function $\Omega=\Omega(\erad,m)$, with $m$ defined in \eqref{eq.defm}:
\begin{itemize}
    \item[(a)] It transforms as $ \xi^t\partial_t\Omega+\xi^x\partial_x\Omega \overset{os}{=}\{\Omega, H_g[\xi^tN^x+\xi^x]+D_g[\xi^tN]\}$ on shell.
    \item[(b)] It commutes with itself, $\{\Omega [s_1],\Omega [s_2]\}=0$.
    \item[(c)] It satisfies the antisymmetry property $\{\Omega [s_1], H_g[s_2]\}+\{H_g[s_1],\Omega [s_2]\}=0$.
\end{itemize}
{It should be emphasized that these properties do not rely on $m$ being constant over solutions, because they are satisfied even outside the constraint surface.}
Properties (a) and (b) are immediate because $\Omega$ is a spacetime scalar that depends on $\qbat'$ but no other derivatives, and it is independent of $\pbat$. To prove property (c), one has to compute the Poisson brackets and remove all derivatives of $s_1$ through integration by parts, 
\begin{align}
    \{\Omega [s_1], H_g[s_2]\}+\{H_g[s_1],\Omega [s_2]\}=-2\int  s_1s_2'\left(\frac{\partial \Omega}{\partial\pbi}\frac{\partial \ham_g}{\partial\qbi'}+\frac{\partial \ham_g}{\partial\pbat}\frac{\partial \Omega}{\partial\qbat'}\right)dx.    
\end{align}
It is a straightforward computation to check that the sum inside brackets on the right-hand side is zero.

By property (a), $\overline{\ham}_g$ follows $\big\{D_g[s_1],\overline{H}_g[s_2]\big\}=\overline{H}_g\big[s_1s_2'\big]$, and properties (b) and (c) ensure that $ \big\{\overline{H}_g[s_1],\overline{H}_g[s_2]\big\}=D_g\big[\overline{F}(s_1s_2'-s_1's_2)\big]$, with $\overline{F}:=\Omega^{-2}F$, if and only if these relations are satisfied by $\ham_g$ and $F$ instead.

Besides, and using again property (a), $\overline{F}$ follows \eqref{eq.covariance} provided $F$ does so, and the correct gauge transformations of the Lagrange multipliers \eqref{gaugelapse} and \eqref{gaugeshift}, are also satisfied if they are followed by $N$ and $N^x$, because the product $\overline{N}^2 \overline{F}=N^2F$ remains unaltered \cite{AlonsoBardaji:2023bww}.

All this shows that such a transformation leaves invariant the phase-space solutions. Nevertheless, the geometry is different, and it acquires an additional conformal factor in the Lorentzian sector,
\begin{align}\label{eq.metricnew}
    ds^2=-{N}^2dt^2+\frac{1}{F}\big(dx+N^xdt\big)^2+r^2d\sigma^2=\Omega^{-2}\left(-\overline{N}^2dt^2+\frac{1}{\overline{F}}\big(dx+N^xdt\big)^2\right)+r^2d\sigma^2.
\end{align}
In spherical symmetry the spacetime manifold is a warped product between the two-sphere and a two-dimensional Lorentzian manifold, and a transformation like the above changes the curvature scalars in a very specific way. Let \eqref{eq.metricnew} be denoted as $ds^2=g_{(2)}+r^2d\sigma^2$, and $d\overline{s}^2=\overline{g}_{(2)}+r^2d\sigma^2$, with ${g}_{(2)}=\Omega^{-2}\overline{g}_{(2)}$. The Ricci scalars are related as
\begin{align}               {\mathcal{R}}&=\Omega^2\left(\overline{\mathcal{R}}+\overline{\nabla}^A\overline{\nabla}_A\log\Omega\right)+\frac{2}{r^2}\big(1-\Omega^2\big),
\end{align}
with $\overline{\nabla}$ the covariant derivative associated with $\overline{g}_{(2)}$. This means that curvature divergences on $d\overline{s}^2$ might be resolved if they coincide with some root of $\Omega$, as explicitly shown in Ref.~\cite{Alonso-Bardaji:2023qgu}.

\end{document}